\documentclass[a4paper, onecolumn]{quantumarticle}
\pdfoutput=1
\usepackage[utf8]{inputenc}
\usepackage[english]{babel}
\usepackage[T1]{fontenc}

\usepackage{adjustbox}
\usepackage{braket}
\usepackage{amssymb}
\usepackage{amsmath}
\usepackage{bm}
\usepackage{mathdots}
\usepackage{graphicx}
\usepackage{mathtools,amssymb,amsmath, amsthm,stmaryrd,thmtools,braket}
\usepackage[numbers,sort&compress]{natbib}
\usepackage{ulem}
\usepackage[ruled,vlined]{algorithm2e}
\usepackage{xr-hyper}
\usepackage[colorlinks=true, urlcolor=blue,citecolor=blue,anchorcolor=blue]{hyperref}
\usepackage[capitalize]{cleveref}

\newcommand{\tinyspace}{\mspace{1mu}}

\newcommand{\norm}[1]{\left\lVert\tinyspace#1\tinyspace\right\rVert}

\renewcommand{\vec}[1]{\bm{#1}}

\usepackage{tikz}
\usetikzlibrary{quantikz}

\DeclareMathOperator{\GL}{GL}
\DeclareMathOperator{\PGL}{PGL}
\DeclareMathOperator{\SO}{SO}

\begin{document}
\title{A New Angle on Quantum Subspace Diagonalization for Quantum Chemistry}
\author{Xeno De Vriendt$^{*}$}
\affiliation{Next Generation Computing, BASF SE, 67061 Ludwigshafen, Germany}
\author{Jacob Bringewatt$^{*}$}
\affiliation{The Volgenau Department of Physics, United States Naval Academy, Annapolis, MD 21402, USA}
\affiliation{Department of Physics, Harvard University, Cambridge, MA 02138, USA}
\author{Nik O. Gjonbalaj}
\affiliation{Department of Physics, Harvard University, Cambridge, MA 02138, USA}
\author{Stefan Ostermann}
\affiliation{Department of Physics, Harvard University, Cambridge, MA 02138, USA}
\author{Davide Vodola}
\affiliation{Next Generation Computing, BASF SE, 67061 Ludwigshafen, Germany}
\author{Johannes Borregaard}
\affiliation{Department of Physics, Harvard University, Cambridge, MA 02138, USA}
\author{Michael Kühn$^{\dagger}$}
\affiliation{Next Generation Computing, BASF SE, 67061 Ludwigshafen, Germany}
\author{Susanne F. Yelin$^{\ddagger}$}
\affiliation{Department of Physics, Harvard University, Cambridge, MA 02138, USA}

\begin{abstract}
Quantum subspace diagonalization and quantum Krylov algorithms offer a feasible, pre- or early-fault tolerant alternative to quantum phase estimation for using quantum computers to estimate the low-lying spectra of quantum systems. However, despite promising proof-of-principle results, such methods suffer from high sensitivity to noise (including intrinsic sources such as sampling noise), making their utility for realistic industry-relevant problems an open question. To improve the potential applicability of such methods, we introduce a new variant of thresholding for noisy generalized eigenvalue problems that arise in quantum subspace diagonalization that has the potential to better control sensitivity to noise. Our approach leverages eigenvector-preserving transformations (rotations) of the generalized eigenvalue problem prior to thresholding. We study this effect in practical settings by applying this rotation thresholding scheme to an iterative quantum Krylov algorithm for several chemical systems, including the industry-relevant Fe(III)-NTA chelate complex. We develop a particular heuristic to select the rotation angle from noisy data and find for certain systems and noise regimes that the samples required to reach a target error for ground state estimation can be reduced by a factor of up to 100. Furthermore, with oracle access to the optimal transformation, more dramatic improvements are possible and we observe reductions in sample requirements by up to $10^4$, motivating the continued development of methods that can realize these improvements in practice. While we develop our approach in the context of quantum subspace diagonalization, the improved thresholding scheme we develop could be advantageous in any context where one must solve noisy, ill-conditioned generalized eigenvalue problems.
\end{abstract}

\maketitle
\def\thefootnote{*}\footnotetext{These authors contributed equally to this work.}\def\thefootnote{\arabic{footnote}}
\def\thefootnote{$\dagger$}\footnotetext{Corresponding author, email: michael.b.kuehn@basf.com.}\def\thefootnote{\arabic{footnote}}
\def\thefootnote{$\ddagger$}\footnotetext{Corresponding author, email: yelin@g.harvard.edu.}\def\thefootnote{\arabic{footnote}}

\section{Introduction}

Computing the low-lying spectrum of quantum systems is a ubiquitous problem in computational physics, from quantum chemistry and condensed matter systems to high-energy physics. Unfortunately, for many systems of interest this problem can be computationally intractable---for instance, when the low lying eigenstates are highly entangled---which makes it a natural and appealing target for quantum simulators and quantum computers. While quantum computers provide a scalable approach via the quantum phase estimation (QPE) algorithm \cite{Cao.2019.Chem.Rev., Mcardle.2020.Rev.Mod.Phys.}, this method requires large-scale fault-tolerant quantum computing hardware \cite{Elfving.2020.Arxiv}, due to the hefty circuit depth requirements \cite{Reiher.2017.Proc.Natl.Acad.Sci.U.S.A., Goings.2022.Proc.Natl.Acad.Sci.U.S.A., Low.2025.Phys.Rev.X}. 

This reality has driven the development of alternative algorithms that require shorter circuit depths and fewer qubits in hopes of obtaining a practical quantum advantage sooner. Of course, reducing the quantum resources comes at a cost: these approaches typically suffer from increased runtime, measurements, or classical post-processing and generally lack rigorous performance guarantees, making their practical utility an important open question~\cite{Bharti.2022.Rev.Mod.Phys.,Tong.2022.Quantum.Views,eisert2025mind}.

Recently, algorithms based on diagonalization within a subspace of basis states, such as quantum subspace diagonalization (QSD) and quantum Krylov diagonalization (QKD), have received significant attention as they have both rigorous performance bounds~\cite{Epperly.2022.SIAM,Kirby.2024.Quantum, Lee.2024.Quantum} and are amenable to proof-of-principle implementations on current devices~\cite{Yoshioka.2025.Nat.Commun., Piccinelli.2025.ArXiv, Nutzel.2025.Quantum.Science}. Additionally, QKD algorithms have provable guarantees of convergence by generating the basis states by the repeated application of the Hamiltonian, acting on the initial state vector, building a so-called Krylov subspace \cite{Golub.2013, Cortes.2022.Phys.Rev.A}. The resulting quantum Krylov diagonalization methods are particularly advantageous as they do not require extensive parameter optimization. In these approaches, one must solve a generalized eigenvalue problem of the form $H_0\vec{\lambda}=\lambda S_0\vec{\lambda}$ using a classical computer, where $H_0$ is the effective Hamiltonian in the subspace and $S_0$ is the overlap matrix of the basis states, respectively. Diagonalizing the Hamiltonian using non-orthogonal bases has a long tradition in quantum chemistry under the label of non-orthogonal configuration interaction (NOCI) \cite{Lowdin.1950.J.Chem.Phys., King.1967.J.Chem.Phys., Koch.1993.Chem.Phys.Lett., Sundstrom.2014.J.Chem.Phys., Mayhall.2014.Phys.Chem.Chem.Phys., Lee.2022.J.Chem.Theory.Comput., DeBaerdemacker.2023.Adv.Quantum.Chem.,  DeVriendt.2023.Mol.Phys.}, and there have been many research efforts directed at finding quantum computational counterparts, such as a non-orthogonal quantum eigensolver \cite{Huggins.2020.New.J.Phys., Baek.2023.PRX.Quantum} and various QSD \cite{Mcclean.2017.Phys.Rev.A, Colless.2018.Phys.Rev.X} and QKD \cite{Motta.2020.Nat.Phys., Parrish.2019.ArXiv, Stair.2020.J.Chem.Theory.Comput., Low.2019.Quantum, Anderson.2025.Quantum, OLeary.2025.Quantum, Byrne.2025.ArXiv, Kanno.2023.ArXiv, Piccinelli.2025.ArXiv, Yang.2025.Phys.Rev.A, Boyd.2025.Phys.Rev.A, Ren.2025.ArXiv} algorithms.

The intuition behind QSD is that there exist sets of non-orthogonal quantum states that are easily preparable on quantum computers, but are not efficiently representable classically, which can be used to define a subspace with much greater support on the target ground state than could be achieved by classical methods.

While this is a plausible argument, QSD has a major issue that classical version does not: the quantum algorithm only has noisy access to the matrix elements of $H_0$ and $S_0$. Furthermore, these matrices are typically ill-conditioned. Thus, quantum subspace diagonalization requires solving a noisy, ill-conditioned generalized eigenvalue problem. Due to the ill-conditioning, we can expect, in general, large perturbations of the generalized eigenvalues due to noise, making the results of these methods unreliable. This issue is especially pronounced for near-to-intermediate term devices---precisely the regime where such methods are expected to be a useful alternative to QPE.

Nonetheless, one can consider various regularization techniques to mitigate these concerns. Standard choices include Tikhonov regularization, where small multiples of the identity are added to the overlap matrix $S_0$ and, sometimes, $H_0$, and thresholding, where one projects both $H_0$ and $S_0$ to the orthogonal complement of the space spanned by the low-lying eigenvectors of $S_0$. While this naive thresholding scheme has a long history as a heuristic numerical technique\footnote{It was first introduced by Löwdin \cite{Lowdin.1967.Rev.Mod.Phys.} under the name of canonical orthogonalization.} it has only recently been rigorously analyzed in the context of quantum subspace diagonalization \cite{Epperly.2022.SIAM,Kirby.2024.Quantum}. 

\begin{figure}[!ht]
\centering\includegraphics[width=\textwidth]{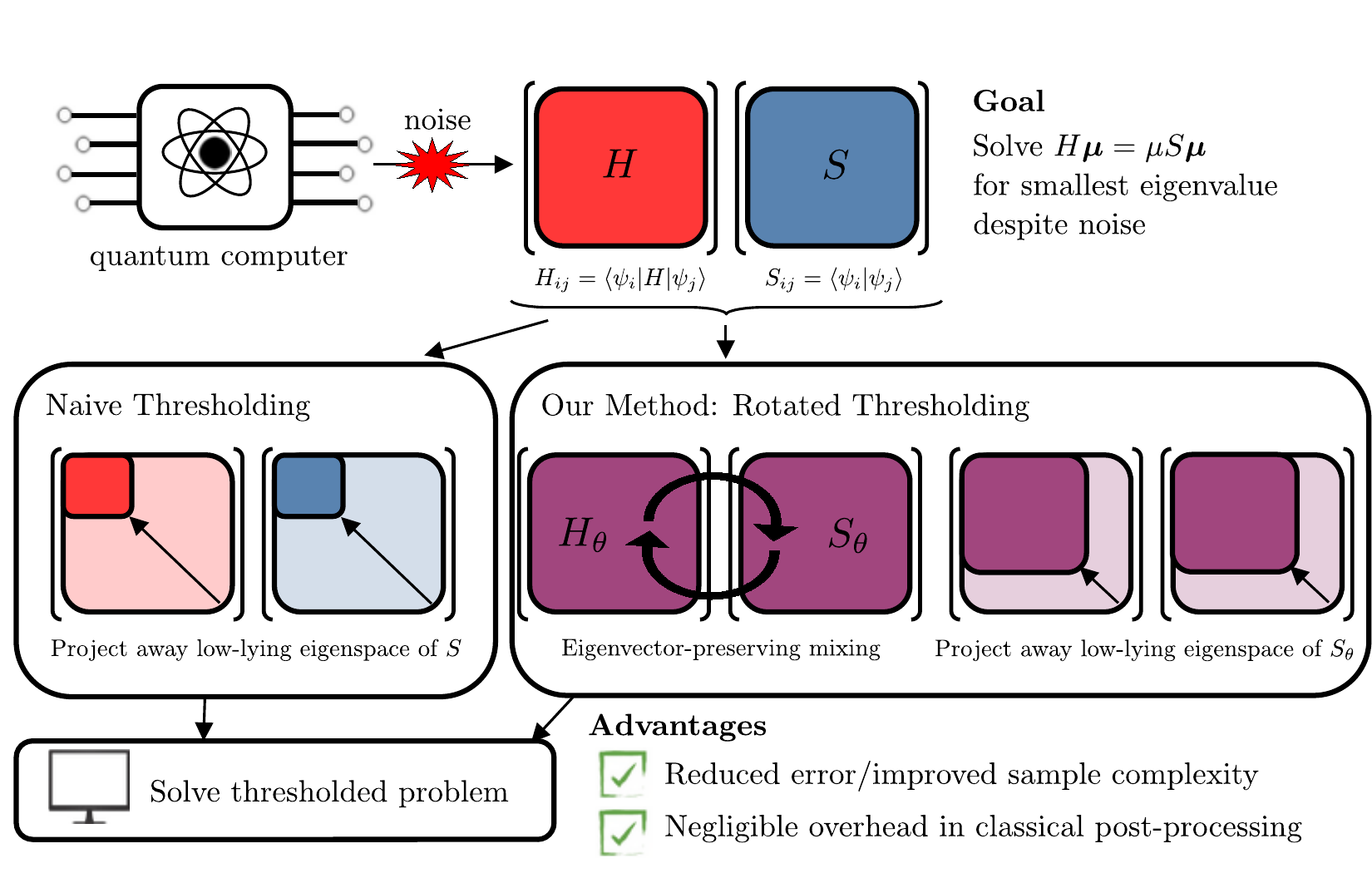}
\caption{A visual overview of naive versus rotation‑based thresholding for quantum subspace diagonalization. After measuring Hamiltonian ($H$) and overlap ($S$) matrix elements on a quantum device, one obtains a noisy and ill‑conditioned generalized eigenvalue problem. To solve it classically typically requires thresholding to project out the low‑lying eigenspace of $S$, where the thresholding parameter that determines what subspace we project away from is determined by the amount of noise. Our rotation approach mixes $H$ and $S$ while preserving their generalized eigenvectors, allowing a less aggressive projection for the same thresholding parameter. Thus, we obtain reduced errors and sampling requirements with minimal classical overhead.}
\label{fig:overview}
\end{figure}

In this paper, we develop an improved variant of thresholding for noisy generalized eigenvalue problems and use this novel computational procedure to show tighter thresholding error bounds for real-time quantum Krylov methods. In particular, we consider applying eigenvector-preserving transformations (rotations) to the generalized eigenvalue problem prior to thresholding. Thresholding breaks the equivalence between the transformed generalized eigenvalue problems, leading to tighter error bounds and, consequently, smaller sampling requirements. This approach is summarized in \cref{fig:overview}.

We show that incorporating this rotation into the thresholding scheme improves the accuracy of ground state energy estimates at a fixed noise level (i.e., with the same number of samples taken from the quantum computer), which can allow us to achieve chemical accuracy at higher noise levels, thereby reducing the sampling requirements. With oracle access to the optimal choice of rotation, we find up to $10^4$ reduction in samples required for certain chemical systems, setting an upper limit on the available improvements for these systems.  With a realistic choice of heuristic to pick the rotation angle the improvements are more minor---up to a factor of about 100---motivating further investigation into heuristics that further close the gap with optimal performance.
In particular, we run our quantum Krylov diagonalization calculations for a series of all-trans polyenes of various lengths and, as a more complicated industry-relevant example, for a Fe(III)-nitrilotriacetate (NTA) chelate complex~\cite{Hehn.2024.J.Chem.Theory.Comput.}. The trans-polyenes are well-known examples of molecules with interesting ground- and excited state structure, and they form the central motifs for a large set of biological compounds \cite{Hu.2015.J.Chem.Theory.Comput., Manna.2020.J.Chem.Phys., Verma.2025.J.Chem.Theory.Comput.}. Furthermore, increasing the number of conjugated bonds in the system gradually increases the multi-reference character \cite{Hu.2015.J.Chem.Theory.Comput.}, thus providing an ideal chemical test-bench for novel method development.

The paper is organized as follows: First, in \cref{s:main-idea} we review the QKD algorithm and describe our improved thresholding scheme in detail. \cref{s:algorithm} describes the full noise-aware Krylov algorithm, and in \cref{s:numerical_results} we apply the full algorithm in combination with the novel thresholding approach to different chemical systems. In \cref{s:conclusion} we summarize our results and discuss possibilities for future work.

\section{Rotated Quantum Subspace Diagonalization}\label{s:main-idea}

\subsection{Quantum Subspace Diagonalization}\label{s:subspacediagonalization}

Let $\hat{H}$ be a Hamiltonian of dimension $N$ with eigenvalues $\lambda_0\leq \lambda_1\leq\cdots\leq \lambda_{N-1}$ and corresponding eigenvectors $\vec{\lambda}_0, \cdots,\vec{\lambda}_{N-1}$. The goal of quantum subspace diagonalization is to estimate the low-lying eigenvalues of $\hat{H}$ via classical diagonalization within a subspace of states prepared on a quantum computer---i.e. via the Rayleigh-Ritz method~\cite{Arfken.2011}.

In particular, consider a set of $d$ basis states $\{\vec{\psi}_j\}_{j=0}^{d-1}$. Letting these basis states form the columns of a matrix
\begin{equation}
V = \left(\vec{\psi}_0,\cdots,\vec{\psi}_{d-1}\right),
\end{equation}
we can define an effective \textit{projected Hamiltonian} $H_0\in\mathbb{C}^{d\times d}$ and an \textit{overlap matrix} $S_0\in\mathbb{C}^{d\times d}$ as
\begin{subequations}
\begin{align}
H_0 &= V^\dagger \hat{H}V \\
S_0 &= V^\dagger V.
\end{align}
\end{subequations}

Under the assumption that the basis states can be efficiently prepared on a quantum computer, one then extracts the estimates of the matrix elements of $H_0$ and $S_0$ via measurements on the quantum device using, e.g., a Hadamard test or multi-fidelity estimation protocol \cite{Cortes.2022.Phys.Rev.A}. Crucially, due to the sampling required to obtain estimates of the matrix elements as well as errors in the basis state preparation, we only obtain access to noisy versions of the true $H_0$ and $S_0$ matrices. Thus, the matrices we consider are
\begin{align}
H=H_0 + E \\
S=S_0 + F,
\end{align}
where $E$ and $F$ are noise matrices, originating from sampling and hardware noise~\cite{Epperly.2022.SIAM}. By sampling just the upper triangular part of $H_0$ and $S_0$ we can enforce that $E$ and $F$ are Hermitian (and thus, so are $H$ and $S$). The magnitude of the noise can be quantified by 
\begin{align}\label{eq:eta}
\eta:= \sqrt{\norm{E}^2+\norm{F}^2},
\end{align}
where $\norm{\cdot}$ is the spectral norm.\footnote{This is an upper bound on the spectral radius $r(E+iF):=\max_{\norm{\vec{x}}=1} \vec{x}^\dagger(E+iF)\vec{x}$. } 

Given the noisy matrix pencil $(H,S)$ one then solves the (noisy) generalized eigenvalue problem
\begin{equation}\label{eq:unrotated-eigenvalue-prob}
H\vec{\mu} = \mu S\vec{\mu},
\end{equation}
on a classical computer for the low-lying (generalized) eigenvalues. We denote these eigenvalues as $\mu_0\leq\mu_1\leq\cdots\leq \mu_{d-1}$ and let the eigenstate associated with the eigenvalue $\mu_j$ be denoted as $\vec{\mu_j}$. The generalized eigenvalues (eigenvectors) obtained are known as Ritz values (vectors). Provided the span of the column space of $V$ has good overlap with the true low-lying states $\vec{\lambda}_j$ and sampling noise is sufficiently small one can expect the Ritz values $\mu_j$ to closely approximate the true eigenvalues $\lambda_j$. This intuition can be formalized in terms of rigorous error bounds~\cite{Epperly.2022.SIAM,Kirby.2024.Quantum}.

In practice, however, the generalized eigenvalue problem in \cref{eq:unrotated-eigenvalue-prob} is ill-conditioned and, for realistic levels of noise, some sort of regularization must be performed before solving. As described in the introduction, the standard choice is what we call ``naive thresholding'', where one projects both $H$ and $S$ away from the eigenspace of $S$ with eigenvalues below some threshold parameter $\tau$ before solving. 

Different choices of basis lead to different variants of quantum subspace diagonalization. For instance, a common choice is to generate the basis states $\vec{\psi_j}$ via real-time evolution so that, for some small time interval $\delta t$
\begin{equation}
\vec{\psi}_j = e^{-i\hat{H}j\delta t}\vec{\psi}_0,
\end{equation}
where $\vec{\psi}_0$ is some ansatz state with reasonable overlap with the low-lying eigenspace of $\hat{H}$. For instance, in the context of quantum chemistry this might be the Hartree-Fock state. By choosing time evolved states to span the subspace, the basis states form a Krylov basis and, as such, ensure that $\mu_j$ converge to the true eigenvalues $\lambda_j$ as the number of basis states increases. This flavor of QKD is known as the real-time quantum Krylov method and its performance, when coupled with naive thresholding is well understood~\cite{Epperly.2022.SIAM,Kirby.2024.Quantum}.

\subsection{From Naive Thresholding to Rotated Thresholding}\label{s:thresholding}
We propose to improve the thresholding step in quantum subspace methods by first transforming the associated generalized eigenvalue problem.It is well-known that the standard eigenvalue problem $A\vec{\alpha}=\alpha \vec{\alpha}$ can be transformed by letting $A\rightarrow \mu A+ \nu I$, for some $\mu,\nu\in\mathbb{R}$, where $I$ is the identity matrix. Such transformations leave the eigenvectors fixed and change the eigenvalues via the same transformation $\alpha \rightarrow \mu\alpha +\nu$. 

This concept of eigenvector-preserving transformations can be naturally extended to generalized eigenvalue problems. In particular, one can consider a symmetrized form of the generalized eigenvalue problem in \cref{eq:unrotated-eigenvalue-prob} as
\begin{equation}
\beta H\vec{\mu} = \alpha S\vec{\mu},
\end{equation}
where the generalized eigenvalue $\mu = \frac{\alpha}{\beta}$. Thus, generalized eigenvalues can be associated with rays $(\alpha,\beta)t\in\mathbb{R}^2$ for $t\in\mathbb{R}^+$, which we call ``eigenrays''. As a consequence, transformations of $\mathbb{R}^2$ that map rays to rays can be viewed as transformations of eigenvalues of generalized eigenvalue problems that leave the eigenvectors invariant. These transformations are elements of the projective general linear group $\PGL_2(\mathbb{R}):=\GL_2(\mathbb{R})/\{\mu I\}$---that is, the group of two-by-two, real invertible matrices $M$ up to a scalar factor.
Transforming the eigenrays of the generalized eigenvalue problem $(H,S)$ via some $M\in \PGL_2(\mathbb{R})$ implies the same transformation on the matrices $(H_0,S_0)$:
\begin{equation}
(H, S) \rightarrow (M_{11} H + M_{21} S, M_{12} H + M_{22} S),
\end{equation}
where $M_{ij}$ are the matrix elemments of the $2\times 2$ matrix $M$.
Observe that if we require $S=I$ both before and after the transformation, we recover a transformation of a standard eigenvalue problem of the form $H\rightarrow \mu H+ \nu I$, as one should expect.

Given that such transformations connect equivalent generalized eigenvalue problems it is natural to ask why such transformations are relevant at all. The crucial idea is that when we consider ill-conditioned, noisy matrix pencils, such as those that arise in quantum Krylov methods, thresholding breaks the equivalence between these generalized eigenvalue problems. Thus, transforming matrix pencils prior to thresholding can lead to better control over the errors.

For this purpose, it is natural to restrict our attention to rotation matrices $M\in \SO(2)\subset \PGL_2(\mathbb{R})$ as rotations do not increase the strength of the perturbations $E$ and $F$ to the matrices $H_0$ and $S_0$.\footnote{This follows from the fact that the numerical radius of the rotated $E$ and $F$ is equal to the numerical radius of the original perturbation matrices.} Thus, we consider the one-parameter family of transformed generalized eigenvalue problems 
\begin{align}\label{eq:rotated-eigenvalue-prob}
\beta_\theta H_\theta\vec{\mu} = \alpha_\theta S_\theta\vec{\mu},
\end{align}
where
\begin{align}
H_\theta = H\cos\theta - S\sin\theta,\nonumber\\
S_\theta = S\cos\theta + H\sin\theta.
\end{align}
The eigenvectors of the rotated problem in \cref{eq:rotated-eigenvalue-prob} are equivalent, up to a global phase, to those in the original unrotated problem \cref{eq:unrotated-eigenvalue-prob}, and the eigenvalues change in a the same way as the matrices:
\begin{align}
\alpha_\theta = \alpha\cos\theta - \beta\sin\theta,\nonumber\\
\beta_\theta = \beta\cos\theta + \alpha\sin\theta.
\end{align}
Thus the original eigenvalues can be recovered simply by reverting of the rotation on the solution pair $(\alpha_\theta,\beta_\theta)=$ such that $\mu_\theta=\frac{\alpha_\theta}{\beta_\theta}$. 

To see how the equivalence between the rotated and unrotated problems can be broken by thresholding it is helpful to consider a toy example. 
In particular, consider a three-dimensional Krylov basis $V=\{\vec{\psi}_0,\vec\psi_1,\vec\psi_2\}$ with associated unrotated overlap matrix
\begin{equation}
S_0 = \begin{pmatrix}
1 & 0 & 0 \\
0 & 1 & s \\
0 & s & 1
\end{pmatrix},
\end{equation}
where $s\in[0,1]$ is a parameter controlling the overlap of the basis elements $\vec\psi_1$ and $\vec\psi_2$. The state $\vec\psi_0$ is orthogonal to the rest of the Krylov space. The eigenvalues of $S_0$ are $\{1-s, 1, 1+s\}$ with corresponding eigenvectors 
\begin{align}
\vec{e}_\pm=\frac{\vec\psi_1\pm\vec\psi_2}{\sqrt{2}}, \qquad \vec{e}_0 =\vec\psi_0. 
\end{align}

For simplicity, suppose the ground state of the full Hamiltonian is within the Krylov subspace so that when there is no noise or thresholding the ground state eigenvalue could be recovered exactly (i.e. there is no Rayleigh-Ritz error due to restricting to a subspace). A particularly simple Hamiltonian is one that is diagonal in the eigenbasis of $S_0$. In particular, suppose 
\begin{align} \label{eq:toy-Hamiltonian}
\hat{H}=H_0 &= \xi \vec{e}_-\vec{e}_-^\dagger +  \left(\frac{\xi}{1-s}+\delta\right) \vec{e}_0\vec{e}_0^\dagger + \left(\frac{1+s}{1-s}\xi+\Delta\right) \vec{e}_+\vec{e}_+^\dagger,
\end{align}
for some choice of constants $\xi,\delta,\Delta>0$. Thus, the eigenvalues of $\hat{H}$ are $\lambda_0 = \frac{\xi}{1-s}$, $\lambda_1 = \lambda_0+\delta$, $\lambda_2=\lambda_1+\Delta$.

\begin{figure}[!hb]
\centering\includegraphics[width=\textwidth]{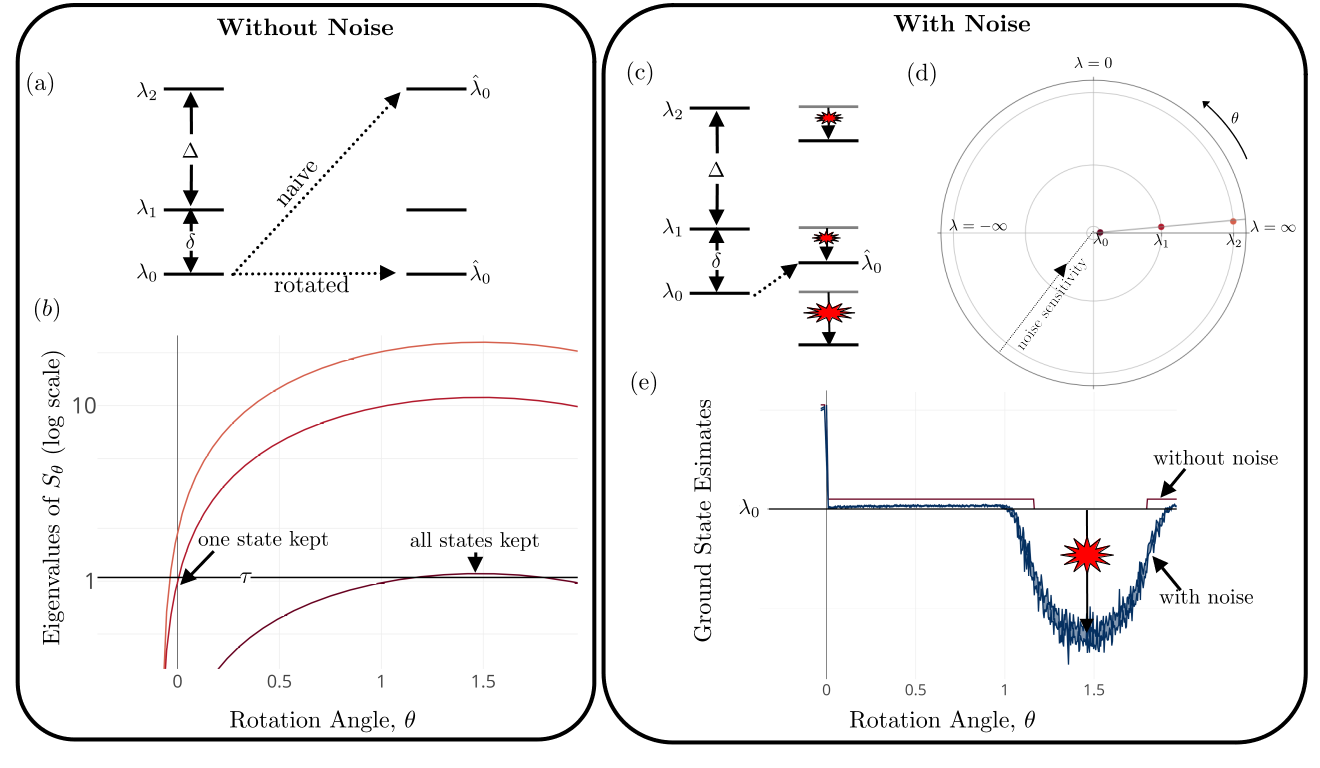}
\caption{For the dimension three example discussed in the main text, panel (a) depicts the eigenvalue structure of the true Hamiltonian and indicates what the estimate $\hat{\lambda}_0$ of the true ground state $\lambda_0$ would be for naive and rotated thresholding (at the optimum rotation angle) without noise. (b) shows the eigenvalues of the rotated overlap matrix $S_\theta$ as a function of rotation angle with a parameter choice $\delta=0.1$, $\Delta=2.0$, $\xi=1.1$, and $s=0.9$. Observe that with no rotation ($\theta=0$) the subspace kept for thresholding at $\tau=1.05$ is only dimension one, whereas for the optimal rotation angle ($\theta\approx 1.4$) all states are kept and the ground state energy is recovered perfectly. (c) depicts the impact of Gaussian noise (chosen with standard deviation of $0.1$) on the estimated ground state energy, indicating that rotated thresholding will pick the perturbed second eigenstate as the best estimate of the ground state. The fact that noise lowers all eigenvalues is an artifact of the simplicity of this model and not a general feature. The fact that one state is still projected away by rotated thresholding when we add noise is explained by panel (d) showing the $(\alpha,\beta)$ plane where rotations act. Perturbation theory for generalized eigenvalue problems~\cite{Epperly.2022.SIAM,Parlett.1998, Mathias.2004.M.C.C.M., Cheng.1999.Lin.Algebra.App.} tells us, roughly speaking, that the closer the eigenvector-normalized coordinates of an eigenvalue are to the origin in this space the more sensitive that eigenvalue is to perturbation. Thus, in this problem the ground state is extremely sensitive to perturbations and is eliminated by thresholding. (e) shows the result of this Gaussian noise on the error estimates as a function. Observe that the best estimate now occurs for $\theta\approx 0.5$ where two states are kept, balancing the impacts of thresholding error and error due to noise.}\label{fig:3by3example}
\end{figure}

Now consider a threshold $\tau$ slightly greater than 1. Naive thresholding on the unrotated problem would throw away the subspace spanned by $\big\{\vec{e}_0,\vec{e}_-\big\}$ as the corresponding eigenvalues are both $\leq$ 1. The ground state $\lambda_0$ would thus be estimated to be $\lambda_2$, which is a poor estimate if $\delta$ or $\Delta$ is large. On the other hand, as shown in \cref{fig:3by3example}a and the top panel of \cref{fig:3by3example}b), for certain parameter choices and with no noise, there exists a rotation angle $\theta$ such that all eigenvalues of the rotated overlap matrix, $S_\theta$ are above threshold. Thus, if we ignore the noise matrices $E$ and $F$, rotated thresholding allows for perfect recovery of the true ground state energy in this problem. The parameters for the results in the figure are $\delta=0.1,$ $\Delta=2.0$, $\xi=1.1$, and $s=0.9$.

\subsection{Optimization of the Rotation Angle}\label{s:angle}
As illustrated starkly by the toy example, in the absence of noise, optimizing the choice of rotation angle to minimize the Rayleigh-Ritz error that arises from projecting away from the low-lying eigenspace of the the overlap matrix $S_\theta$ is intuitively straightforward: simply choose a rotation angle such that thresholding at some value $\tau$ discards the smallest possible subspace. Indeed, a rigorous bound on the error incurred due to thresholding alone is known (see Thm. 4.2 of Ref.~\cite{Epperly.2022.SIAM}), which applies equally well to the rotated version of the generalized eigenvalue problem as the unrotated one. In particular, the results of Thm. 4.2 of  \cite{Epperly.2022.SIAM} suggest that to minimize the error due to thresholding we should pick our rotation precisely to minimize $\norm{c_\theta}$ where $c_\theta$ is the eigenvector associated with the smallest eigenvalue of the generalized eigenvalue problem, normalized so that $c_\theta^\dagger S_\theta c_\theta=1$. 

While a rigorously understood condition for minimizing thresholding error, in practice, this is not a viable cost function since it requires knowledge of the true eigenvector $c_\theta$. But there is a bigger issue: we cannot simply choose a rotation that minimizes the error due to thresholding, as the impact of the perturbations $E$ and $F$ also changes with rotation angle and actual problems have noise. Indeed, if there is no noise there is no reason to apply a threshold at all and any reductions in thresholding error due to a rotation is a moot point.

The toy example described above nicely demonstrates the trade-offs inherent in choosing a rotation angle. As shown in \cref{fig:3by3example}(c)-(e), when we add Hermitian, Gaussian noise on the matrix elements of $S_0$ and $H_0$ with standard deviation $\delta$, one finds that the optimal angle for estimating the ground state energy is no longer the one that keeps the most states after thresholding. Rather, an optimal rotation angle retains a subspace of dimension two, balancing out minimizing the error due to thresholding and the error due to sample noise. The intuition behind this result is that if we keep a noisy state that is not particularly important for obtaining a good estimate of the ground state, the addition of this noise may outweigh the advantages obtained from lowering the thresholding error by keeping as many states as possible.

\section{Detailed Implementation of the Rotated Quantum Krylov Algorithm}\label{s:algorithm}
Having established the main ideas, in this section, we discuss in detail the implementation of the rotated quantum Krylov method, up to the existence of an oracle for the choice of optimal rotation angle. While oracle access establishes the ability of rotations to improve thresholding, we will, in \cref{s:heuristics}, make a first step towards a practical heuristic that does not require oracle access. Future work could focus on the importance of this oracle and improved heuristics.

Beyond the basic scheme described in the previous section, our implementation of a rotated quantum Krylov algorithm introduces several additional features to manage the noise in the matrix elements. 
First, certain properties of the true matrices $H_0$ and $S_0$, like the fact they are Hermitian or, in the case of $S_0$, positive semi-definite, can be violated in the presence of noise. Since we know our matrices should have these properties we can enforce them. For instance, Hermiticity of the measured Hamiltonian and overlap matrices can be ensured by only measuring the upper triangle of off-diagonal matrix elements (or only the first row for Hermitian Toeplitz matrices \cite{Byrne.2025.ArXiv}). 

Ensuring that the noisy overlap matrix $S$ is positive semi-definite is more challenging, as we cannot achieve this through similar measurement shortcuts. While this problem is usually circumvented by artificially setting any negative eigenvalues of the overlap matrix to zero \cite{Kirby.2024.Quantum}, and thus discarding problematic parts of the subspace through thresholding, the eigenvector preserving rotation of the matrix pencil requires access to the positive semi-definite matrix $S$ before thresholding. Thus, we find the closest matrix $X$ to $S$, which adheres to the constraints of a physical overlap matrix, by minimizing the Frobenius norm of the difference between these two matrices, 
\begin{equation}
    \min_{X} \| X - S \|_F \thinspace ,
\end{equation}
under the constraints of physicality
\begin{equation} \label{eq:pos-def}
    \begin{cases}
        &X = X^\dagger \\
        &X \succeq 0 \\
        &\text{diag}(X) = 1
    \end{cases} \thinspace ,
\end{equation}
i.e. Hermitian, positive semi-definite, and all diagonal elements equal to one. This is a convex optimization problem that is generally easy to solve. Because the true overlap matrix satisfies these constraints, the projection will---at worst---maintain the existing noise level.

Another major issue when dealing with noisy matrix elements is that noise breaks the variationality of the quantum Krylov algorithm. As such, incorporating too much noise in the matrix representations can yield non-physical energies. To account for this behavior, we derive a noise-aware convergence criterion for the subspace dimension, leveraging Gaussian statistics derived from our measurements. 

We directly measure the matrix elements $m$ separate times on the quantum computer, resulting in $m$ noisy matrix pencils $(H_i, S_i$, $\forall i \in [1, m])$. We then divide these $m$ different measurements into $n$ separate batches consisting of $p$ Hamiltonian or overlap matrices, respectively. We construct a \textit{batch average} of both $H$ and $S$ $\forall q \in [1, n]$ as
\begin{subequations}
    \begin{align}
        \bar{H}_q &= \frac{1}{n} \sum_{i = (q-1)p}^{qp - 1} H_i \thinspace ,\\
        \bar{S}_q &= \frac{1}{n} \sum_{i = (q-1)p}^{qp - 1} S_i \thinspace .
    \end{align}
\end{subequations}
    
For each batch average matrix pencil $(\bar{H}_q, \bar{S}_q)$, we solve the generalized eigenvalue problem to find the batch average eigenenergies $\vec{\bar{\mu}}_q$, with $\bar{\mu}_{0, q}$ the ground state energy. \textit{Between} different $\bar{\mu}_{0, q}$, we subsequently calculate the global weighted average
\begin{equation} \label{eq:batch-avrg-E}
    \bar{\mu}_{0} = \frac{1}{n} \sum_{q=1}^{n} w_q \bar{\mu}_{0, q} \thinspace ,
\end{equation}
where the weight $w_q$ of each entry is determined by the number of measurement shots in each batch (e.g. all $1$ for a constant batch size $p$), and the associated standard deviation
\begin{equation} \label{eq:std-dev}
    \sigma = \sqrt{\frac{1}{n} \sum_{q=1}^{n}(\bar{\mu}_{0, q} - \bar{\mu}_{0})^2} \thinspace ,
\end{equation}
which we square to find the variance $\sigma^2$. To control the iterative growth of the Krylov subspace, we now define the noise-aware convergence criterion as the largest standard deviation (\cref{eq:std-dev}) of the previous two iterations $j$, scaled by a factor $\gamma$ 
\begin{equation}
    \varepsilon = \gamma \max(\sigma_j, \sigma_{j-1}) \thinspace .
\end{equation}

When the difference between two subsequent $\bar{\mu}_{0}$ (\cref{eq:batch-avrg-E}) is smaller than $\varepsilon$, the last Krylov state added to the many-body basis causes a statistically irrelevant energy shift, likely due to noise, and the iterative procedure stops. If the change in energy between iterations falls well below chemical accuracy (i.e. is smaller than $10^{-4}$), the iterative procedure also stops. The basis construction procedure is described in \cref{alg:convergence}. 

Once the basis dimension has converged, we build an aggregate matrix pencil $(\bar{H}_m, \bar{S}_m)$ by taking the weighted average of all separate $\bar{H}_q$ and $\bar{S}_q$, 
\begin{subequations}
    \begin{align}
        \bar{H}_m &= \frac{1}{n} \sum_{q=1}^{n} w_q \bar{H}_q \thinspace , \label{eq:global-avrg-H} \\
        \bar{S}_m &= \frac{1}{n} \sum_{q=1}^{n} w_q \bar{S}_q \thinspace , \label{eq:global-avrg-S}
    \end{align}
\end{subequations}
from the complete set of $m$ measurements. Solving the generalized eigenvalue problem associated with this aggregate matrix pencil yields the final eigenenergies and eigenstates. If naive thresholding would cause a collapse in subspace dimension while solving this generalized eigenvalue problem, we invoke the rotation thresholding scheme (see \cref{s:thresholding}) in a bid to minimize the thresholding error while simultaneously avoiding the introduction of excess noise (see \cref{s:angle}).

Assuming oracle access to the optimal rotation angle, we rotate the global weighted average matrix pencil (\cref{eq:global-avrg-H}, \cref{eq:global-avrg-S}) and solve the rotated problem. After the necessary back-transformations this yields the low lying energy states in the Krylov many-body subspace, as well as the expansions of those states in the defined Krylov basis.

\begin{algorithm}[H]
    \caption{Noise aware iterative basis construction, starting after adding at least $1$ additional state to the Krylov subspace.} \label{alg:convergence}
    $\mathcal{K} = \{ \psi_0, \psi_1 \}$
    \For{$j = 2$ \KwTo \texttt{max\_iterations}}{
        $\psi_j = e^{-i \hat{H} j \delta t} \psi_0$\;
        $\mathcal{K}$.append($\psi_j$)\;
        Measure and batch up matrices: $\bar{H}_q$ and $\bar{S}_q$ $\forall q$\;
        Find $\bar{X}_q$:  $\min_{\bar{X}_q} \| \bar{X}_q - \bar{S}_q \|_F \thinspace \forall q$\;
        Solve $\bar{H}_q C_q = \vec{\bar{\mu}}_q \bar{X}_q C_q$ $\forall q$, calculate $\bar{\mu}_{0, j}$, $\sigma_j$ and $\varepsilon_j$\;
        \If{$|\bar{\mu}_{0, j} - \bar{\mu}_{0, j-1}| < \varepsilon_j$ or $|\bar{\mu}_{0, j} - \bar{\mu}_{0, j-1}| < 10^{-4}$}{
            \textbf{break}\;
        }
    }
    \Return $\bar{H}_q$ and $\bar{S}_q$ $\forall q$ \;
\end{algorithm}

\section{Numerical Results for Chemical Systems}\label{s:numerical_results}

\begin{figure}[ht!]
    \centering
    \includegraphics[width=\linewidth]{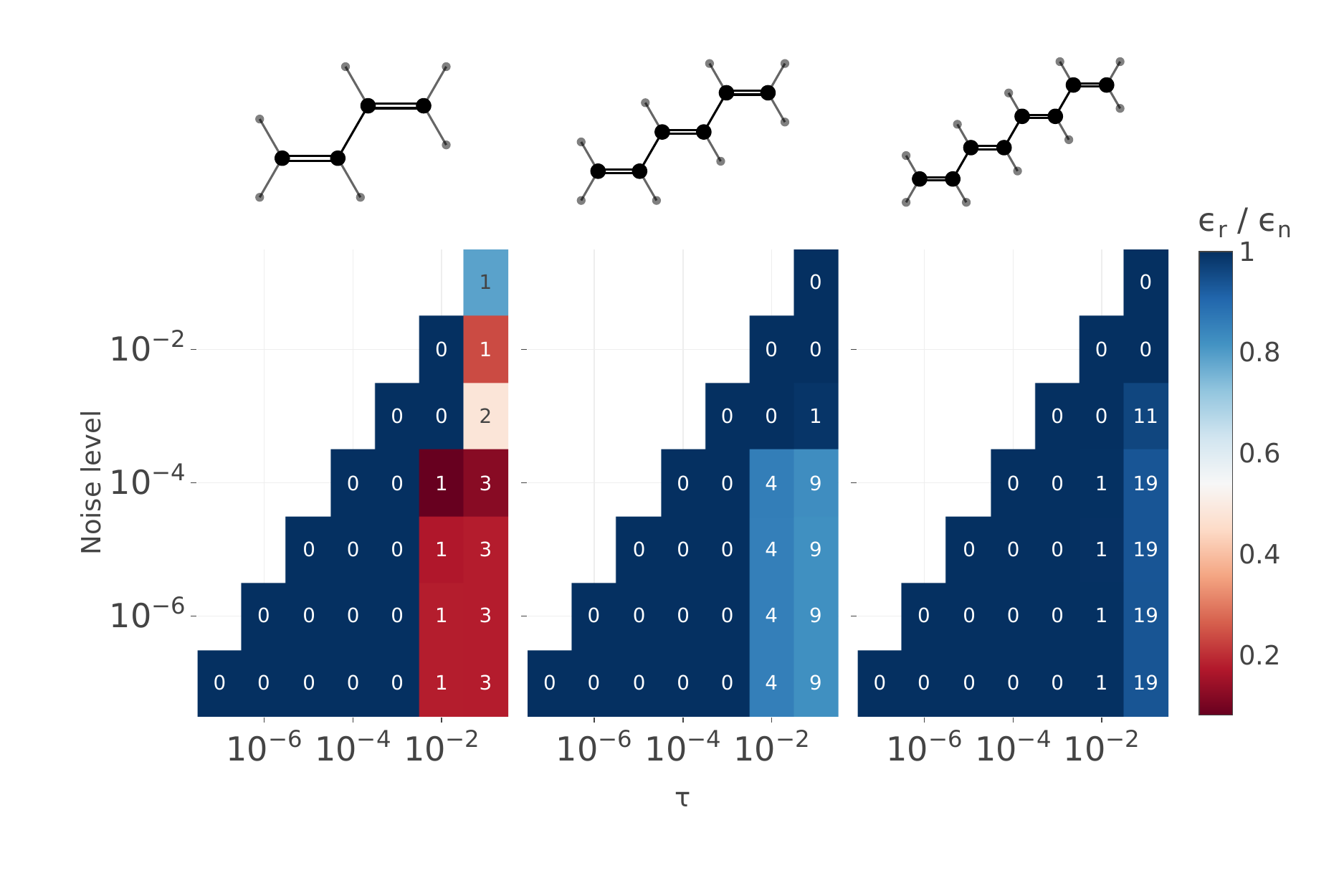}
    \caption{Ratio of the absolute error with respect to the CASSCF solution made by the rotation thresholding approach with rotation angle $\theta_{\text{optimal}}$ ($\epsilon_r$) and the same error made by naively thresholding ($\epsilon_n$), calculated for in length increasing all-trans polyene chains. Noise level indicates the variance of the artificially applied Gaussian noise, and $\tau$ is the thresholding parameter. Numbers in the heatmap represent the amount of additional states kept by the rotation approach, that were discarded when applying naive thresholding. Convergence scaling factor $\gamma=1.0$.}
    \label{fig:polyenes-exact-grid}
\end{figure}

\begin{figure}[ht!]
    \centering
    \includegraphics[width=\linewidth]{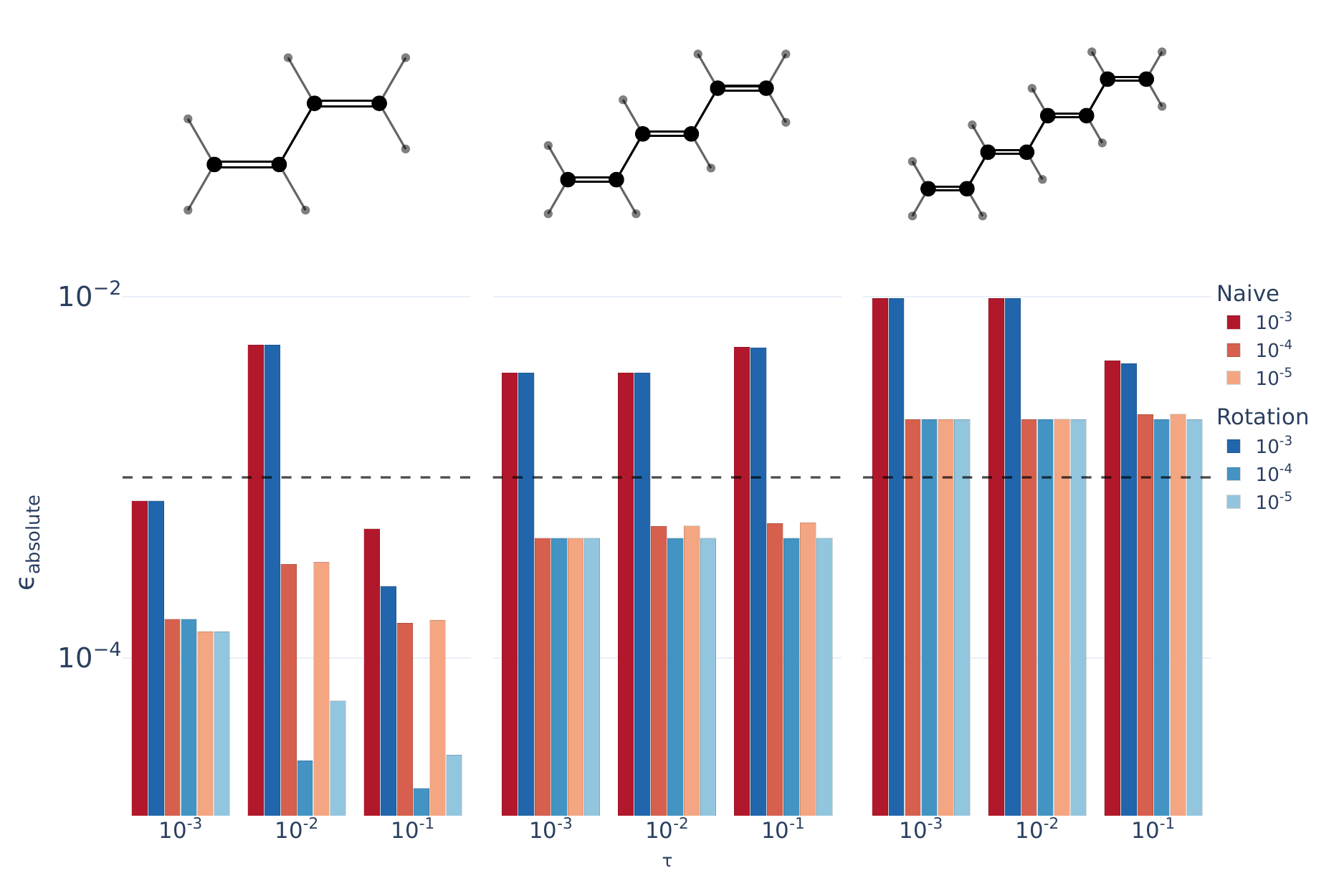}
    \caption{Absolute error ($\epsilon_\text{absolute}$) in log-scale with respect to the CASSCF solution made by the rotation thresholding approach with rotation angle $\theta_{\text{optimal}}$ (blue bars) and the same error made by naively thresholding (red bars) for in length increasing all-trans polyene chains and three different noise levels, at all thresholding parameters $\tau \geq$ noise level. Dashed line represents chemical accuracy. Convergence scaling factor $\gamma=1.0$.}
    \label{fig:polyenes-error-bars}
\end{figure}

In this section, we will show that rotation thresholding, in combination with a noise aware construction of the real-time quantum Krylov basis, leads to either equal or improved accuracy for actual chemical systems compared to the quantum Krylov algorithm with naive thresholding. For our test cases, we select the time step for the construction of the real-time Krylov basis based on the spectral gap of the Hamiltonian ($\delta t = \frac{\pi}{\Delta E}$), calculated using the complete active space self-consistent field (CASSCF) method. We simulate the noise resulting from measuring the matrix elements of the generalized eigenvalue problem by adding a Gaussian noise matrix, of which we can tune the variance to increase the severity of the noise, i.e. the noise level. Gaussian noise is a good approximation for sampling noise in the limit of many measurements \cite{Lee.2024.Quantum}. Provided the sampling noise is the dominant noise source, we take the simulated noise as a proxy for \textit{all} noise. We simulate the measurement of 100 noisy batches, each consisting of 1250 noise instances at the chosen noise level. To find the closest positive semi-definite overlap matrix to the noisy one, we solve the convex optimization problem from \cref{eq:pos-def} using the Cvxpy package \cite{Diamond.2016.J.Mach.Learn.Res.}. We assume oracle access to the optimal rotation angle, $\theta_{\text{optimal}}$. In practice, the optimization of $\theta_{\text{optimal}}$ is carried out by minimizing the energy difference of the respective ground states calculated with the rotated Krylov method and the CASSCF method. In \cref{s:heuristics}, we discuss a potential optimization heuristic to avoid the need for an accurate approximation of the true ground state energy. In both cases, we minimize the scalar function using a line search algorithm taken from the Scipy suite of programs \cite{Virtanen.2020.Nat.Methods}. CASSCF references are calculated using PySCF \cite{Sun.2020.J.Chem.Phys.}. The corresponding active space Hamiltonians are generated using OpenFermion \cite{McClean.2020.Quantum.Sci.Technol.}.
For all the following results, we only take into account the data points where the noise level is equal to or lower than the chosen thresholding parameter, as one would do in a practical experiment.

\subsection{All-trans Polyenes}\label{s:polyenes}

To test the performance of the rotated thresholding approach with scaling active spaces and gradually increasing levels of correlation, we study a series of all-trans polyenes, gradually increasing the number of conjugated bonds from two (butadiene) to four (octatetraene). For each $\pi$-system, we define the $\pi$-$\pi^*$ orbitals as the active space \cite{Verma.2025.J.Chem.Theory.Comput.}. As such, each conjugated bond is treated as a (2, 2) active space. We define the geometry of these systems using the following semi-empirical values \cite{Roberts.1965, Hoffmann.1966.Tetrahedron}: $r_{C-C} = 1.46$Å, $r_{C=C} = 1.34$Å, $r_{C-H} = 1.09$Å and the ideal bond angle of $120^\circ$. For all polyenes, the Krylov basis is constructed iteratively with a maximum of 34 states added on top of the Hartree-Fock state. Our polyene calculations are carried out in the cc-pVDZ basis set.

We study a grid of different noise levels and thresholding parameter ($\tau$) values, and calculate the ratio of the errors made by our rotated thresholding approach and naive thresholding (\cref{fig:polyenes-exact-grid}). The grid points where this ratio is smaller than one show that rotation thresholding outperforms its naive counterpart in almost half of the grid points for butadiene, by discarding between one and three fewer Krylov basis states than the naive approach (see numbers in the grid of \cref{fig:polyenes-exact-grid}). Due to the effects of noise, naive thresholding causes Krylov basis states that provide a relevant contribution to the physical description of the system to be discarded. The rotation thresholding approach increases the stability of these states and, by keeping states that have \textit{only} become unstable due to noise effects, reduces the thresholding error. 

\cref{fig:polyenes-error-bars} shows the absolute energy errors made by naive and rotation thresholding respectively, for combinations of noise level and thresholding parameter inside the application domain extracted from \cref{fig:polyenes-exact-grid}. At a noise level of $10^{-4}$ and thresholding parameters $\tau = 10^{-1}$ and $\tau = 10^{-2}$, the rotation-improved Krylov method gives a much lower absolute error than the best result achieved with the naive thresholding approach (noise level of $10^{-5}$, $\tau = 10^{-3}$). Where rotation thresholding decreases the errors at noise level $10^{-4}$ to $0.019$ and $0.027$ millihartree for $\tau = 10^{-1}$ and $\tau = 10^{-2}$ respectively, the best result that naive thresholding is able to achieve still has an absolute error of $0.14$ millihartree, meaning that the rotation reduces the error by $\pm$80\% to $\pm$86\% (\cref{fig:polyenes-error-bars}). Because the sampling noise scales inversely with the square root of the number of samples ($\text{noise}\propto\frac{1}{\sqrt{\text{samples}}}$), being able to achieve the same quality of results ($\epsilon_{\text{absolute, naive}} = 0.14$ millihartree at $\tau=10^{-3}$, $\epsilon_{\text{absolute, rotation}} = 0.25$ millihartree, at $\tau=10^{-1}$) at a noise level that is two orders of magnitude higher ($10^{-3}$ for the rotation case, $10^{-5}$ for the naive case) means that up to a factor of $10^4$ fewer samples can be taken from the quantum computer without significantly reducing the quality of the final result. Furthermore, we can improve on the best naive thresholding result (noise level $10^{-5}$ and $\tau = 10^{-3}$) by taking $10^2$ fewer samples when applying the rotation (noise level $10^{-4}$ and $\tau = 10^{-1}$).

Although the region where rotation thresholding outperforms naive thresholding becomes smaller for the larger polyene chains, the rotation approach still leads to the lowest absolute error for a fixed noise level across all thresholding parameters, as evidenced by the absolute rotation error (the blue bar) always being smaller or, at worst, equal in height to the absolute error made by naive thresholding (the corresponding red bar) (\cref{fig:polyenes-error-bars}). However, the difference between the rotation errors and the naive errors is generally small for hexatriene and octatetraene. To see whether these seemingly diminishing returns are related to parameters in the Krylov algorithm (e.g. maximum number of Krylov basis states, convergence scaling $\gamma$, \dots) or are actually related to the increase in system size and correlation, we expand on our chemical test-bench with the industry-relevant Fe(III)-NTA chelate complex.

\begin{figure}[ht!]
    \centering
    \includegraphics[width=0.4\linewidth]{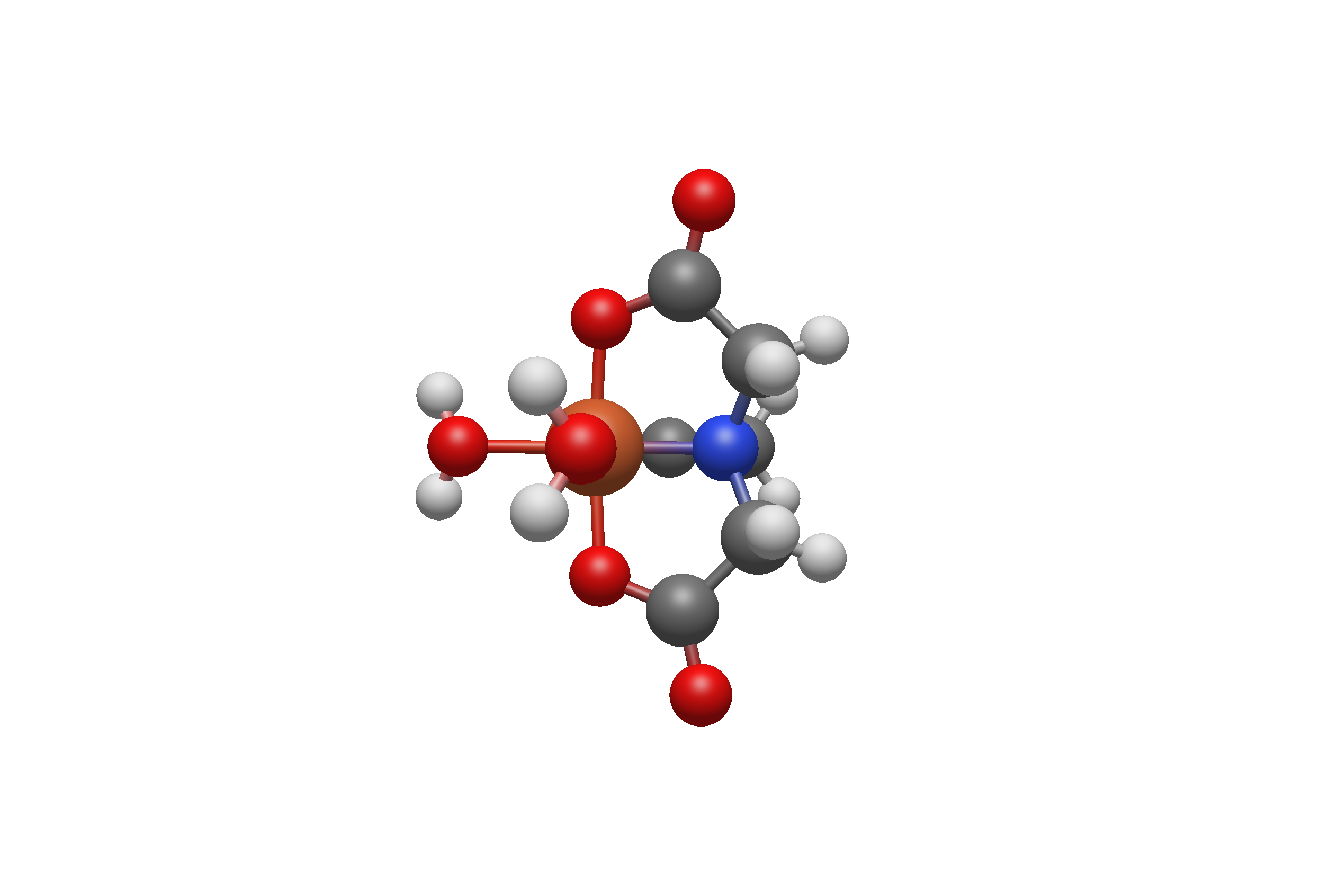}
    \caption{Molecular structure of the low-spin state of the Fe(III)-NTA chelate complex. Hydrogen atoms are shown in light gray, carbon atoms in dark gray, oxygen atoms in red, nitrogen atoms in blue and Fe(III) is shown in orange.}
    \label{fig:FeNTA-structure}
\end{figure}
\begin{figure}[ht!]
    \centering
    \includegraphics[width=0.8\linewidth]{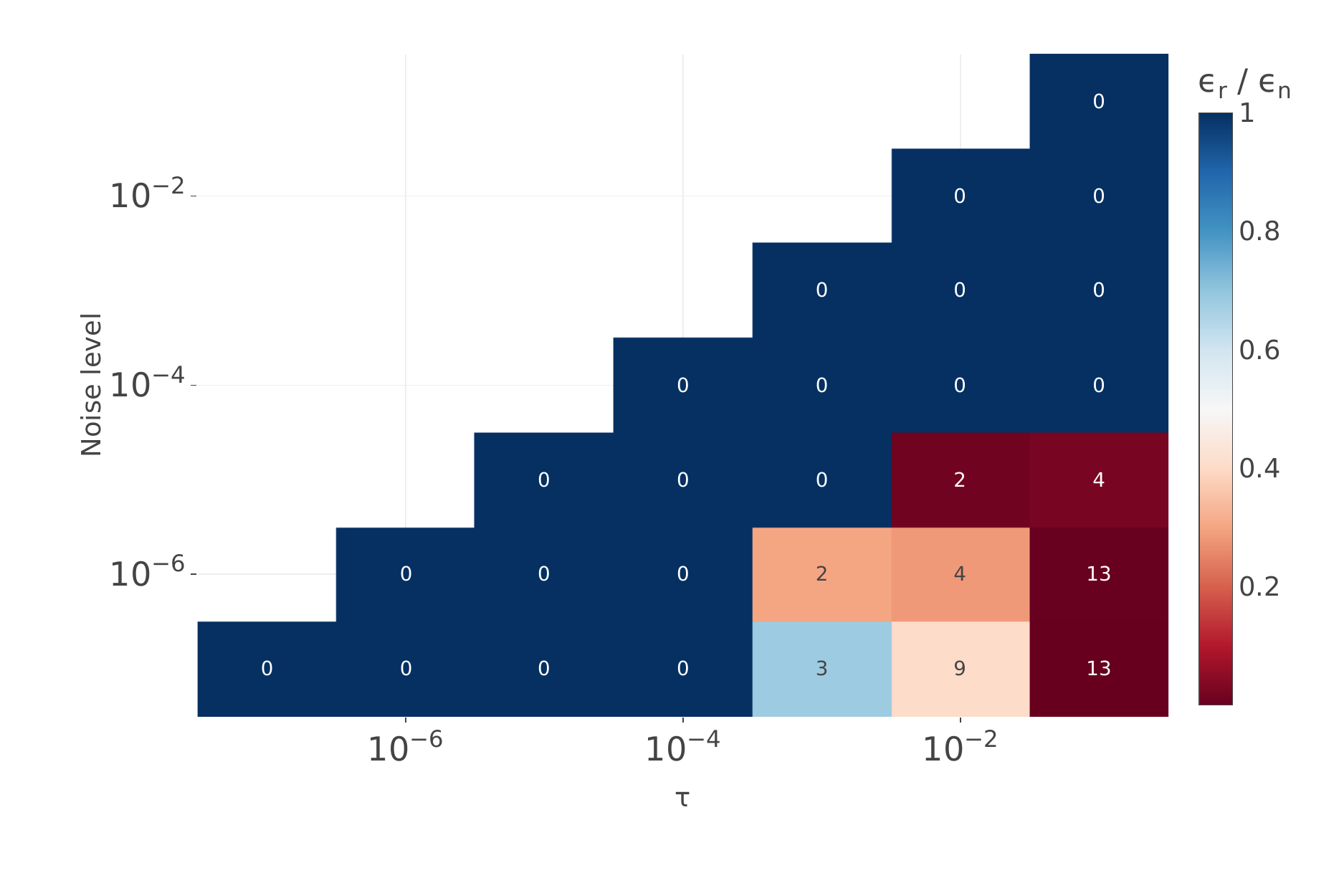}
    \caption{Ratio of the absolute error with respect to the CASSCF solution made by the rotation thresholding approach with rotation angle $\theta_{\text{optimal}}$ ($\epsilon_r$) and the same error made by naively thresholding ($\epsilon_n$), calculated for the Fe(III)-NTA chelate complex. Noise level indicates the variance of the artificially applied Gaussian noise, and $\tau$ is the thresholding parameter. Numbers in the heatmap represent the amount of additional states kept by the rotation approach, that were discarded when applying naive thresholding. Convergence scaling factor $\gamma=0.75$.}
    \label{fig:FeNTA-exact-grid}
\end{figure}
\begin{figure}[ht!]
    \centering
    \includegraphics[width=0.8\linewidth]{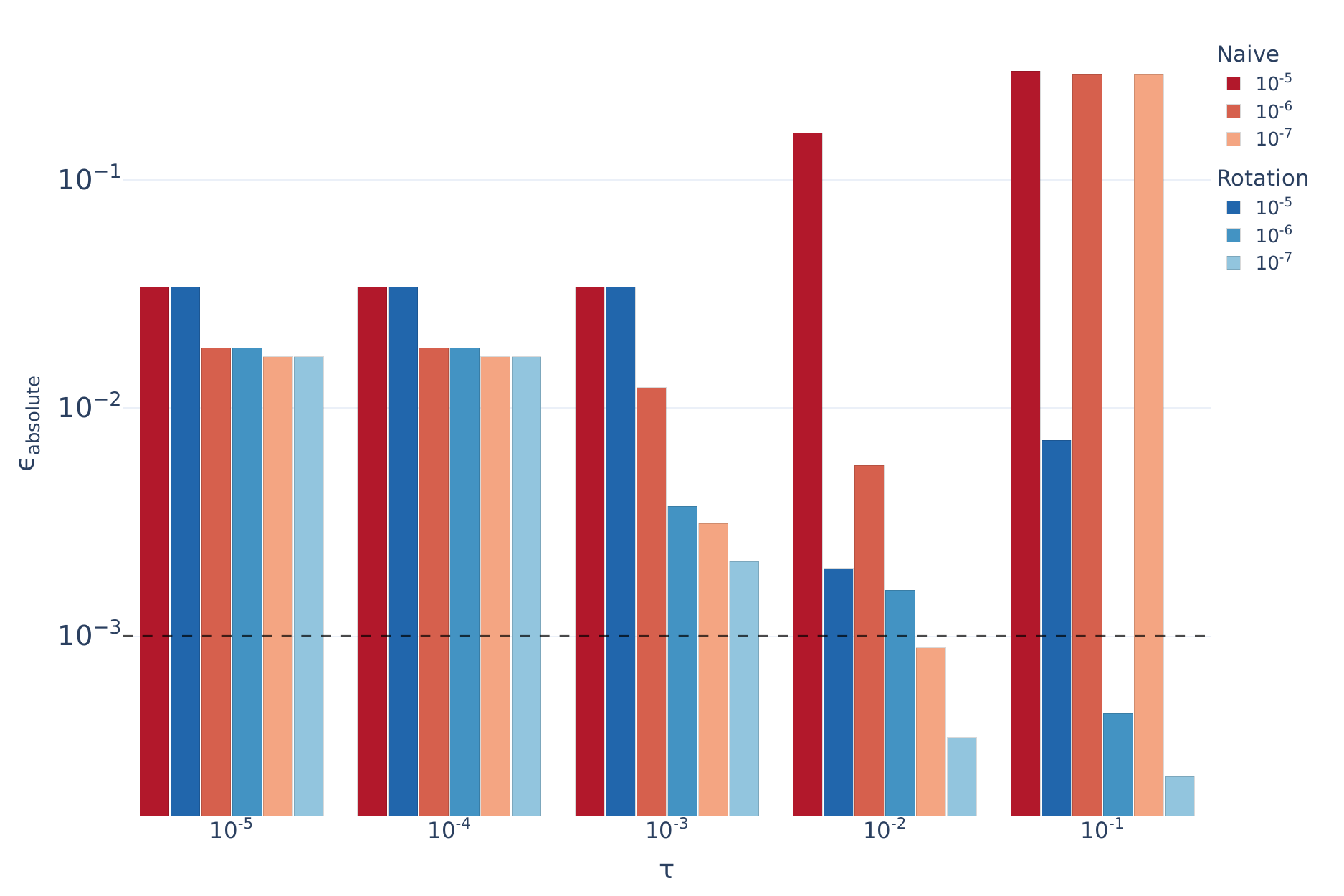}
    \caption{Absolute error ($\epsilon_\text{absolute}$) in log-scale with respect to the CASSCF solution made by the rotation thresholding approach with rotation angle $\theta_{\text{optimal}}$ (blue bars) and the same error made by naively thresholding (red bars) for three different noise levels, at all thresholding parameters $\tau \geq$ noise level, calculated for the Fe(III)-NTA chelate complex. Dashed line represents chemical accuracy. Convergence scaling factor $\gamma=0.75$.}
    \label{fig:FeNTA-error-bars}
\end{figure}

\subsection{Fe(III)-NTA}\label{s:fenta}

The Fe(III)-NTA chelate complex (\cref{fig:FeNTA-structure}) has been extensively studied with state-of-the-art classical methods \cite{Hehn.2024.J.Chem.Theory.Comput.}, and the calculation of its spin-state energetics has emerged as a potential candidate that may benefit from the integration of quantum computing in an industrial workflow, motivating recent proof-of-concept studies on quantum hardware \cite{Nutzel.2025.Quantum.Science}. For this work, we focus on the most strongly correlated spin state of this system, its low-spin state, defined in the (5, 5) active space consisting of five electrons in the five 3d orbitals of the Fe(III)-ion. The geometry is the optimized ground state geometry of the low-spin state, taken from the work of Hehn et al. \cite{Hehn.2024.J.Chem.Theory.Comput.}. The Krylov basis is constructed iteratively with a maximum of 76 states added on top of the Hartree-Fock state. Fe(III)-NTA calculations are run in the def2-QZVPP basis set.

In \cref{fig:FeNTA-exact-grid} we plot the same error ratio grid, this time for the more complex Fe(III)-NTA system. We see a performance increase when the rotation is applied before thresholding in almost 1/3 of the data points, defining a clear applicability domain, similar to butadiene. The absolute errors of both thresholding approaches applied to the Fe(III)-NTA Krylov calculations (\cref{fig:FeNTA-error-bars}) show that rotation thresholding consistently yields the smallest absolute error for all values of $\tau$ and therefore achieves the lowest overall error. Applying rotation thresholding leads to an absolute error reduction of more than 99\% compared to the naive thresholding result (at noise level $10^{-7}$ and $\tau=10^{-1}$), and a reduction of more than 97.5\% at a higher noise level ($10^{-5}$) and several thresholding parameters ($\tau=10^{-2}$ and $\tau=10^{-1}$) (see \cref{fig:FeNTA-error-bars}). Additionally, the absolute error made by rotation thresholding at noise level $10^{-6}$ and $\tau=10^{-1}$ is smaller than the error made by naive thresholding at every noise level and $\tau$. This shows that the rotation thresholding approach can yield ground state energies within chemical accuracy, at a noise level that is one order of magnitude larger than the noise level that would be required to reach chemical accuracy using the naive approach (noise level $10^{-7}$, $\tau=10^{-2}$). Even then, the error made by the rotation approach is still only $\pm$51\% of the absolute naive error. As such, using this rotation reduces the measurement requirements from the quantum computer \textit{and} improves on the approximation of the ground state energy while doing so.

\subsection{Heuristics for $\theta$-optimization} \label{s:heuristics}

So far, we have always assumed oracle access to the optimal rotation angle $\theta_{\text{optimal}}$. In numerics, we obtain this optimal angle via an optimization with the CASSCF solution as a reference. In realistic settings, this oracle is not available, however. Thus, to avoid the need for this reference energy and to develop intuition for the performance of our algorithm subject to realistic assumptions, we introduce a possible heuristic for the rotation angle, $\theta_{\text{heuristic}}$. The heuristic aims to minimize the variance ($\sigma^2$, \cref{eq:std-dev}) across different measurement batches, as described in \cref{alg:theta-opt}. Intuitively, reducing this variance means that the results from each batch become more consistent with each other. As such, by optimizing the rotation angle for this cost function, we effectively balance how many states are discarded during the thresholding procedure in each measurement batch. This in turn prevents outliers based on thresholding errors, leading to a more accurate estimate of the ground state energy $\bar{\mu}_{0}$. We emphasize that this heuristic is simply one reasonable choice. Developing and testing other heuristics remains an important area for further investigation.

\begin{algorithm}[H]
    \caption{Optimization of rotation angle $\theta$ by minimizing $\sigma^2(\theta)$.} \label{alg:theta-opt}
    \For{$\theta_i \in [0, \pi]$}{
        Find $\bar{X}_q$:  $\min_{\bar{X}_q} \| \bar{X}_q - \bar{S}_q \|_F \thinspace \forall q$\;
        Rotate the matrix pencils: ($\bar{H}_q(\theta_i)$, $\bar{X}_q(\theta_i)$) $\forall q$ \;
        Solve $\bar{H}_q(\theta_i) C_q(\theta_i) = \vec{\bar{\mu}}_q(\theta_i) \bar{X}_q(\theta_i) C_{q}(\theta_i)$ $\forall q$ \;
        Calculate $\sigma^2(\theta_i)$\;
        \If{$\sigma^2(\theta_i) < \sigma^2(\theta_{i-1})$}{
            $\theta_{\text{heuristic}} = \theta_i$\;
        }
    }
    \Return $\theta_{\text{heuristic}}$ \;
\end{algorithm}
\begin{figure}[ht]
    \centering
    \includegraphics[width=0.95\linewidth]{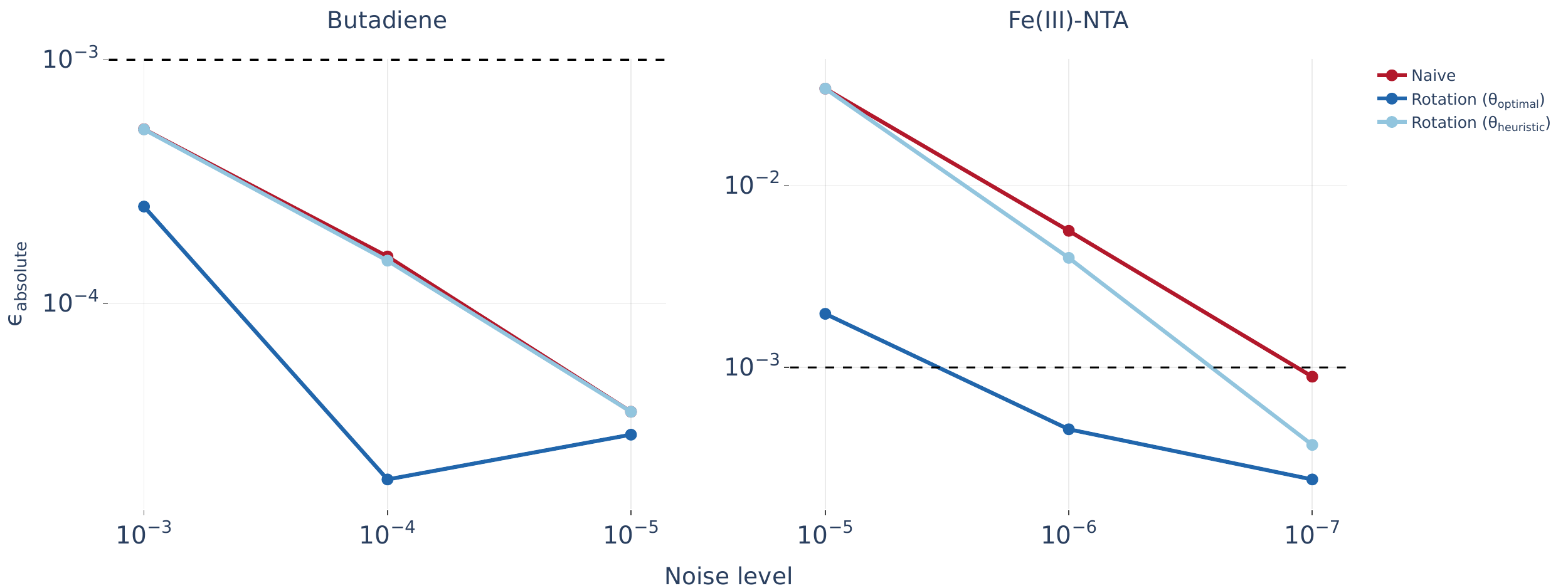}
    \caption{Lowest achieved absolute error with respect to the CASSCF solution across all values of $\tau$ made by naive thresholding (red), rotation thresholding with the optimal rotation angle $\theta_{\text{optimal}}$ (dark blue) and rotation thresholding with the heuristically optimized rotation angle $\theta_{\text{heuristic}}$ (light blue), with respect to the applied noise level for both butadiene (left) and Fe(III)-NTA (right). Dashed line represents chemical accuracy.}
    \label{fig:Errors-wrt-samples}
\end{figure} 

Using the heuristically optimized rotation angle $\theta_{\text{heuristic}}$, we calculated the application domains and absolute errors for the polyene chains and the Fe(III)-NTA chelate complex (See \cref{s:appendix-figures}) and summarized the most important insights from them. We plot the lowest absolute error across thresholding parameters with respect to the relevant noise levels extracted from their respective application domains in \cref{fig:Errors-wrt-samples}. The results for the larger polyene chains (hexatriene and octatetraene) do not change compared to those obtained with the optimal rotation angle $\theta_{\text{optimal}}$. We hypothesize that the reason for this similarity lies in the fact that, given the Krylov parameters (i.e. number of states added, $\delta t$, basis set, convergence scaling $\gamma$, etc.), the rotation is not able to have a larger impact than what we see here. As such we focus on butadiene and Fe(III)-NTA. While we clearly see a possible sampling advantage for butadiene (\cref{fig:Errors-wrt-samples}, left) when using the optimal rotation angle, as evidenced by reaching similar or better accuracy at higher noise levels, the improvements achieved through the heuristic are much smaller. For Fe(III)-NTA, however, rotation thresholding using $\theta_{\text{heuristic}}$ significantly improves upon the naive thresholding results. Even though we do not see the sampling advantage achievable when using $\theta_{\text{optimal}}$, the naive thresholding method allows us to barely reach chemical accuracy ($\epsilon_{\text{absolute}}=0.89$ millihartree) under an applied noise level of $10^{-7}$, while rotation thresholding using $\theta_{\text{heuristic}}$ reduces this error to less than half ($\epsilon_{\text{absolute}}=0.38$ millihartree). Depending on the variance of the results from an actual experiment, this improvement could already be the difference between reaching or not reaching chemical accuracy with certainty.

\section{Conclusion and Outlook}\label{s:conclusion}

In this work, we showed that applying eigenvector preserving transformations to generalized eigenvalue problems prior to thresholding enables tighter control of ill-conditioned, noisy instances of those problems. These improvements are a valuable addition to the classical post-processing step of quantum subspace diagonalization, including quantum Krylov methods. 
Applied to specific chemical systems, rotated thresholding not only shows an improved accuracy for a fixed number of samples compared to the naive thresholding approach but also the potential to achieve better results at noise levels of up to two orders of magnitude larger, or, equivalently, with up to a factor of $10^4$ fewer samples taken from the quantum computer, assuming oracle access to the optimal rotation angle.
Of course, the full extent of these improvements is not accessible in realistic settings due to the reliance on oracle access to the optimal rotation angle. Thus, we also introduced a potential heuristic to optimize the rotation angle. Although the heuristic continues to outperform naive thresholding, the reduction in errors is considerably smaller compared to the optimal angle. Consequently, improving this heuristic or developing alternative ones is a crucial area for further study if the rotated thresholding approach is to provide dramatic improvements for real-world applications. Despite this need, even in situations when the sampling advantage is relatively small, due to the very small classical overhead of the rotation step, our improved thresholding technique is straightforward to incorporate into any quantum subspace algorithm.
The associated reduction in sampling requirements is an important (although, not sufficient) precondition for industry-relevant applications of quantum Krylov methods on near-term quantum hardware. Relatedly, a proof of how much information an oracle for the optimal rotation angle provides regarding the true ground state energy of a system is an interesting target to better understand the extent to which we can expect heuristics to approach the optimal performance.

These QSD methods also yield approximations for excited states. As such, it would be interesting to see how an improvement on the thresholding improves results when one is targeting higher lying energy states instead of focusing on the ground state.

Finally, there are other more sophisticated methods for thresholding, such as the Fix-Heiberger algorithm~\cite{Fix.1972.SIAM.Numer.Anal.}. It would be worthwhile to explore how transformations of the generalized eigenvalue problem of the sort we consider here might interact with these schemes for controlling unstable generalized eigenvalue problems.

\section{Acknowledgments}

We thank Nishad Maskara for many helpful comments on the first version of this manuscript and many interesting discussions regarding potential applications and open questions. X.D.V. gratefully acknowledges C. Ufrecht for many enlightening ad-hoc chats and discussions on both the theory and the results presented in this work. J. Bringewatt notes that the views expressed in this work are those of the author and do not reflect the official policy or position of the United States Naval Academy or any department of the United States Government. S.F.Y. thanks the NSF through the CUA (PHY-2317134) and the Welcome Leap Foundation. 

\bibliography{main}

@article{Epperly.2022.SIAM,
  title={A theory of quantum subspace diagonalization},
  author={Epperly, Ethan N and Lin, Lin and Nakatsukasa, Yuji},
  journal={SIAM Journal on Matrix Analysis and Applications},
  volume={43},
  number={3},
  pages={1263--1290},
  year={2022},
  publisher={SIAM},
  url={https://doi.org/10.1137/21M145954X}
}

@article{Fix.1972.SIAM.Numer.Anal.,
  title={An algorithm for the ill-conditioned generalized eigenvalue problem},
  author={Fix, George and Heiberger, Richard},
  journal={SIAM Journal on Numerical Analysis},
  volume={9},
  number={1},
  pages={78--88},
  year={1972},
  publisher={SIAM},
  url={https://doi.org/10.1137/0709009}
}

@article{eisert2025mind,
  title={Mind the gaps: The fraught road to quantum advantage},
  author={Eisert, Jens and Preskill, John},
  journal={arXiv preprint arXiv:2510.19928},
  year={2025},
  url={https://doi.org/10.48550/arXiv.2510.19928}
}

@article{Cheng.1999.Lin.Algebra.App.,
  title={The nearest definite pair for the Hermitian generalized eigenvalue problem},
  author={Cheng, Sheung Hun and Higham, Nicholas J},
  journal={Linear Algebra and Its Applications},
  volume={302},
  pages={63--76},
  year={1999},
  publisher={Elsevier},
  url={https://doi.org/10.1016/S0024-3795(99)00026-9}
}

@article{Lee.2024.Quantum,
  title={Sampling error analysis in quantum Krylov subspace diagonalization},
  author={Lee, Gwonhak and Lee, Dongkeun and Huh, Joonsuk},
  journal={Quantum},
  volume={8},
  pages={1477},
  year={2024},
  publisher={Verein zur F{\"o}rderung des Open Access Publizierens in den Quantenwissenschaften},
  url={https://doi.org/10.22331/q-2024-09-19-1477}
}

@article{Kirby.2024.Quantum,
  title={Analysis of quantum Krylov algorithms with errors},
  author={Kirby, William},
  journal={Quantum},
  volume={8},
  pages={1457},
  year={2024},
  publisher={Verein zur F{\"o}rderung des Open Access Publizierens in den Quantenwissenschaften},
  url={https://doi.org/10.22331/q-2024-08-29-1457}
}

@article{Bharti.2022.Rev.Mod.Phys.,
  title={Noisy intermediate-scale quantum algorithms},
  author={Bharti, Kishor and Cervera-Lierta, Alba and Kyaw, Thi Ha and Haug, Tobias and Alperin-Lea, Sumner and Anand, Abhinav and Degroote, Matthias and Heimonen, Hermanni and Kottmann, Jakob S and Menke, Tim and others},
  journal={Rev. Mod. Phys.},
  volume={94},
  number={1},
  pages={015004},
  year={2022},
  publisher={APS},
  url={https://doi.org/10.1103/RevModPhys.94.015004}
}

@article{Cao.2019.Chem.Rev.,
  title={Quantum chemistry in the age of quantum computing},
  author={Cao, Yudong and Romero, Jonathan and Olson, Jonathan P and Degroote, Matthias and Johnson, Peter D and Kieferov{\'a}, M{\'a}ria and Kivlichan, Ian D and Menke, Tim and Peropadre, Borja and Sawaya, Nicolas PD and others},
  journal={Chemical reviews},
  volume={119},
  number={19},
  pages={10856--10915},
  year={2019},
  publisher={ACS Publications},
  url={https://doi.org/10.1021/acs.chemrev.8b00803}
}

@article{Mcardle.2020.Rev.Mod.Phys.,
  title={Quantum computational chemistry},
  author={McArdle, Sam and Endo, Suguru and Aspuru-Guzik, Al{\'a}n and Benjamin, Simon C and Yuan, Xiao},
  journal={Rev. Mod. Phys.},
  volume={92},
  number={1},
  pages={015003},
  year={2020},
  publisher={APS},
  url={https://doi.org/10.1103/RevModPhys.92.015003}
}

@article{Elfving.2020.Arxiv,
  title={How will quantum computers provide an industrially relevant computational advantage in quantum chemistry?},
  author={Elfving, Vincent E and Broer, Benno W and Webber, Mark and Gavartin, Jacob and Halls, Mathew D and Lorton, K Patrick and Bochevarov, A},
  journal={arXiv preprint arXiv:2009.12472},
  year={2020},
  url={https://arxiv.org/abs/2009.12472}
}

@article{Nutzel.2025.Quantum.Science,
  title={Solving an industrially relevant quantum chemistry problem on quantum hardware},
  author={N{\"u}tzel, Ludwig and Gresch, Alexander and Hehn, Lukas and Marti, Lucas and Freund, Robert and Steiner, Alex and Marciniak, Christian D and Eckstein, Timo and Stockinger, Nina and Wolf, Stefan and others},
  journal={Quantum Science and Technology},
  volume={10},
  number={1},
  pages={015066},
  year={2025},
  publisher={IOP Publishing},
  url={https://doi.org/10.1088/2058-9565/ad9ed3}
}

@article{Hehn.2024.J.Chem.Theory.Comput.,
  title={Chelate Complexes of 3d Transition Metal Ions─ A Challenge for Electronic-Structure Methods?},
  author={Hehn, Lukas and Deglmann, Peter and Kühn, Michael},
  journal={J. Chem. Theory Comput.},
  volume={20},
  number={11},
  pages={4545--4568},
  year={2024},
  publisher={ACS Publications},
  url={https://doi.org/10.1021/acs.jctc.3c01375}
}

@article{Cortes.2022.Phys.Rev.A,
  title={Quantum Krylov subspace algorithms for ground-and excited-state energy estimation},
  author={Cortes, Cristian L and Gray, Stephen K},
  journal={Phys. Rev. A},
  volume={105},
  number={2},
  pages={022417},
  year={2022},
  publisher={APS},
  url={https://doi.org/10.1103/PhysRevA.105.022417}
}

@book{Arfken.2011,
  title={Mathematical methods for physicists: a comprehensive guide},
  author={Arfken, George B and Weber, Hans J and Harris, Frank E},
  year={2011},
  publisher={Academic press},
  url={https://www.sciencedirect.com/book/monograph/9780123846549/mathematical-methods-for-physicists}
}

@article{Mathias.2004.M.C.C.M.,
  title={The definite generalized eigenvalue problem: A new perturbation theory},
  author={Mathias, Roy and Li, Chi-Kwong},
  journal={Manchester Centre for Computational Mathematics},
  pages={NAREP--457},
  year={2004},
  url={https://www.semanticscholar.org/paper/The-Definite-Generalized-Eigenvalue-Problem-%3A-A-New-Mathias-Li/db81b1f885042dadfc50ea24626abbcb04a2297d}
}

@article{Tong.2022.Quantum.Views,
  title={Designing algorithms for estimating ground state properties on early fault-tolerant quantum computers},
  author={Tong, Yu},
  journal={Quantum Views},
  volume={6},
  pages={65},
  year={2022},
  publisher={Verein zur F{\"o}rderung des Open Access Publizierens in den Quantenwissenschaften},
  url={	https://doi.org/10.22331/qv-2022-07-22-65}
}

@article{Lowdin.1950.J.Chem.Phys.,
  title={On the non-orthogonality problem connected with the use of atomic wave functions in the theory of molecules and crystals},
  author={L{\"o}wdin, Per-Olov},
  journal={J. Chem. Phys.},
  volume={18},
  number={3},
  pages={365--375},
  year={1950},
  publisher={American Institute of Physics},
  url={https://doi.org/10.1063/1.1747632}
}

@article{King.1967.J.Chem.Phys.,
  title={Corresponding orbitals and the nonorthogonality problem in molecular quantum mechanics},
  author={King, Harry F and Stanton, Richard E and Kim, Hojing and Wyatt, Robert E and Parr, Robert G},
  journal={J. Chem. Phys.},
  volume={47},
  number={6},
  pages={1936--1941},
  year={1967},
  publisher={American Institute of Physics},
  url={https://doi.org/10.1063/1.1712221}
}

@article{Koch.1993.Chem.Phys.Lett.,
  title={Linear superposition of optimized non-orthogonal Slater determinants for singlet states},
  author={Koch, Henrik and Dalgaard, Esper},
  journal={Chem. Phys. Lett.},
  volume={212},
  number={1-2},
  pages={193--200},
  year={1993},
  publisher={Elsevier},
  url={https://doi.org/10.1016/0009-2614(93)87129-Q}
}

@article{Sundstrom.2014.J.Chem.Phys.,
  title={Non-orthogonal configuration interaction for the calculation of multielectron excited states},
  author={Sundstrom, Eric J and Head-Gordon, Martin},
  journal={J. Chem. Phys.},
  volume={140},
  number={11},
  year={2014},
  publisher={AIP Publishing},
  url={https://doi.org/10.1063/1.4868120}
}

@article{Mayhall.2014.Phys.Chem.Chem.Phys.,
  title={Spin--flip non-orthogonal configuration interaction: a variational and almost black-box method for describing strongly correlated molecules},
  author={Mayhall, Nicholas J and Horn, Paul R and Sundstrom, Eric J and Head-Gordon, Martin},
  journal={Phys. Chem. Chem. Phys.},
  volume={16},
  number={41},
  pages={22694--22705},
  year={2014},
  publisher={Royal Society of Chemistry},
  url={https://doi.org/10.1039/C4CP02818J}
}

@article{Lee.2022.J.Chem.Theory.Comput.,
  title={Localized Spin Rotations: A Size-Consistent Approach to Nonorthogonal Configuration Interaction},
  author={Lee, Nicholas and Thom, Alex JW},
  journal={J. Chem. Theory Comput.},
  volume={18},
  number={2},
  pages={710--722},
  year={2022},
  publisher={ACS Publications},
  url={https://doi.org/10.1021/acs.jctc.1c00862}
}

@incollection{DeBaerdemacker.2023.Adv.Quantum.Chem.,
  title={Spin-constrained Hartree--Fock and the generator coordinate method for the 2-site Hubbard model},
  author={De Baerdemacker, Stijn and Ayati, Amir and Burton, Hugh GA and De Vriendt, Xeno and Bultinck, Patrick and Acke, Guillaume},
  booktitle={Adv. Quantum Chem.},
  volume={88},
  pages={161--182},
  year={2023},
  publisher={Elsevier},
  url={https://doi.org/10.1016/bs.aiq.2023.03.014}
}

@article{DeVriendt.2023.Mol.Phys.,
  title={Capturing correlation in the spin frustrated H3-ring using the generator coordinate method and spin-constrained generalised Hartree-Fock states},
  author={De Vriendt, Xeno and De Vos, John and De Baerdemacker, Stijn and Bultinck, Patrick and Acke, Guillaume},
  journal={Mol. Phys.},
  volume={121},
  number={9-10},
  pages={e2134831},
  year={2023},
  publisher={Taylor \& Francis},
  url={https://doi.org/10.1080/00268976.2022.2134831}
}

@article{Huggins.2020.New.J.Phys.,
  title={A non-orthogonal variational quantum eigensolver},
  author={Huggins, William J and Lee, Joonho and Baek, Unpil and O’Gorman, Bryan and Whaley, K Birgitta},
  journal={New J. Phys.},
  volume={22},
  number={7},
  pages={073009},
  year={2020},
  publisher={IOP Publishing},
  url={https://doi.org/10.1088/1367-2630/ab867b}
}

@article{Baek.2023.PRX.Quantum,
  title={Say no to optimization: A nonorthogonal quantum eigensolver},
  author={Baek, Unpil and Hait, Diptarka and Shee, James and Leimkuhler, Oskar and Huggins, William J and Stetina, Torin F and Head-Gordon, Martin and Whaley, K Birgitta},
  journal={PRX Quantum},
  volume={4},
  number={3},
  pages={030307},
  year={2023},
  publisher={APS},
  url={https://doi.org/10.1103/PRXQuantum.4.030307}
}

@article{Mcclean.2017.Phys.Rev.A,
  title={Hybrid quantum-classical hierarchy for mitigation of decoherence and determination of excited states},
  author={McClean, Jarrod R and Kimchi-Schwartz, Mollie E and Carter, Jonathan and De Jong, Wibe A},
  journal={Phys. Rev. A},
  volume={95},
  number={4},
  pages={042308},
  year={2017},
  publisher={APS},
  url={https://doi.org/10.1103/PhysRevA.95.042308}
}

@article{Colless.2018.Phys.Rev.X,
  title={Computation of molecular spectra on a quantum processor with an error-resilient algorithm},
  author={Colless, James I and Ramasesh, Vinay V and Dahlen, Dar and Blok, Machiel S and Kimchi-Schwartz, Mollie E and McClean, Jarrod R and Carter, Jonathan and de Jong, Wibe A and Siddiqi, Irfan},
  journal={Phys. Rev. X},
  volume={8},
  number={1},
  pages={011021},
  year={2018},
  publisher={APS},
  url={https://doi.org/10.1103/PhysRevX.8.011021}
}

@book{Golub.2013,
  title={Matrix computations},
  author={Golub, Gene H and Van Loan, Charles F},
  year={2013},
  publisher={JHU press},
  url={https://epubs.siam.org/doi/book/10.1137/1.9781421407944}
}

@book{Parlett.1998,
  title={The symmetric eigenvalue problem},
  author={Parlett, Beresford N},
  year={1998},
  publisher={SIAM},
  url={https://epubs.siam.org/doi/book/10.1137/1.9781611971163}
}

@article{Motta.2020.Nat.Phys.,
  title={Determining eigenstates and thermal states on a quantum computer using quantum imaginary time evolution},
  author={Motta, Mario and Sun, Chong and Tan, Adrian TK and O’Rourke, Matthew J and Ye, Erika and Minnich, Austin J and Brandao, Fernando GSL and Chan, Garnet Kin-Lic},
  journal={Nat. Phys.},
  volume={16},
  number={2},
  pages={205--210},
  year={2020},
  publisher={Nature Publishing Group UK London},
  url={https://doi.org/10.1038/s41567-019-0704-4}
}

@article{Parrish.2019.ArXiv,
  title={Quantum filter diagonalization: Quantum eigendecomposition without full quantum phase estimation},
  author={Parrish, Robert M and McMahon, Peter L},
  journal={arXiv preprint arXiv:1909.08925},
  year={2019},
  url={https://arxiv.org/abs/1909.08925}
}

@article{Stair.2020.J.Chem.Theory.Comput.,
  title={A multireference quantum Krylov algorithm for strongly correlated electrons},
  author={Stair, Nicholas H and Huang, Renke and Evangelista, Francesco A},
  journal={J. Chem. Theory. Comput.},
  volume={16},
  number={4},
  pages={2236--2245},
  year={2020},
  publisher={ACS Publications},
  url={https://doi.org/10.1021/acs.jctc.9b01125}
}

@article{Low.2019.Quantum,
  title={Hamiltonian simulation by qubitization},
  author={Low, Guang Hao and Chuang, Isaac L},
  journal={Quantum},
  volume={3},
  pages={163},
  year={2019},
  publisher={Verein zur F{\"o}rderung des Open Access Publizierens in den Quantenwissenschaften},
  url={	https://doi.org/10.22331/q-2019-07-12-163}
}

@article{Anderson.2025.Quantum,
  title={Solving lattice gauge theories using the quantum Krylov algorithm and qubitization},
  author={Anderson, Lewis W and Kiffner, Martin and O'leary, Tom and Crain, Jason and Jaksch, Dieter},
  journal={Quantum},
  volume={9},
  pages={1669},
  year={2025},
  publisher={Verein zur F{\"o}rderung des Open Access Publizierens in den Quantenwissenschaften},
  url={	https://doi.org/10.22331/q-2025-03-25-1669}
}

@article{OLeary.2025.Quantum,
  title={Partitioned quantum subspace expansion},
  author={O'Leary, Tom and Anderson, Lewis W and Jaksch, Dieter and Kiffner, Martin},
  journal={Quantum},
  volume={9},
  pages={1726},
  year={2025},
  publisher={Verein zur F{\"o}rderung des Open Access Publizierens in den Quantenwissenschaften},
  url={	https://doi.org/10.22331/q-2025-05-05-1726}
}

@article{Byrne.2025.ArXiv,
  title={A Quantum-Centric Super-Krylov Diagonalization Method},
  author={Byrne, Adam and Kirby, William and Soodhalter, Kirk M and Zhuk, Sergiy},
  journal={arXiv preprint arXiv:2412.17289},
  year={2025},
  url={https://arxiv.org/abs/2412.17289}
}

@article{Kanno.2023.ArXiv,
  title={Quantum-selected configuration interaction: Classical diagonalization of Hamiltonians in subspaces selected by quantum computers},
  author={Kanno, Keita and Kohda, Masaya and Imai, Ryosuke and Koh, Sho and Mitarai, Kosuke and Mizukami, Wataru and Nakagawa, Yuya O},
  journal={arXiv preprint arXiv:2302.11320},
  year={2023},
  url={https://arxiv.org/abs/2302.11320}
}

@article{Piccinelli.2025.ArXiv,
  title={Quantum chemistry with provable convergence via randomized sample-based quantum diagonalization},
  author={Piccinelli, Samuele and Baiardi, Alberto and Rossmannek, Max and Vazquez, Almudena Carrera and Tacchino, Francesco and Mensa, Stefano and Altamura, Edoardo and Alavi, Ali and Motta, Mario and Robledo-Moreno, Javier and others},
  journal={arXiv preprint arXiv:2508.02578},
  year={2025},
  url={https://arxiv.org/abs/2508.02578}
}

@article{Yang.2025.Phys.Rev.A,
  title={Shadow-based quantum subspace algorithm for the nuclear shell model},
  author={Yang, Ruyu and Wang, Tianren and Lu, Bing-Nan and Li, Ying and Xu, Xiaosi},
  journal={Phys. Rev. A},
  volume={111},
  number={1},
  pages={012620},
  year={2025},
  publisher={APS},
  url={https://doi.org/10.1103/PhysRevA.111.012620}
}

@article{Boyd.2025.Phys.Rev.A,
  title={High-dimensional subspace expansion using classical shadows},
  author={Boyd, Gregory and Koczor, B{\'a}lint and Cai, Zhenyu},
  journal={Phys. Rev. A},
  volume={111},
  number={2},
  pages={022423},
  year={2025},
  publisher={APS},
  url={https://doi.org/10.1103/PhysRevA.111.022423}
}

@article{Ren.2025.ArXiv,
  title={An Error Mitigated Non-Orthogonal Quantum Eigensolver via Shadow Tomography},
  author={Ren, Hang and Zhang, Yipei and Billings, Wendy M and Tomann, Rebecca and Tkachenko, Nikolay V and Head-Gordon, Martin and Whaley, K Birgitta},
  journal={arXiv preprint arXiv:2504.16008},
  year={2025},
  url={https://arxiv.org/abs/2504.16008}
}

@article{Yoshioka.2025.Nat.Commun.,
  title={Krylov diagonalization of large many-body Hamiltonians on a quantum processor},
  author={Yoshioka, Nobuyuki and Amico, Mirko and Kirby, William and Jurcevic, Petar and Dutt, Arkopal and Fuller, Bryce and Garion, Shelly and Haas, Holger and Hamamura, Ikko and Ivrii, Alexander and others},
  journal={Nature Communications},
  volume={16},
  number={1},
  pages={5014},
  year={2025},
  publisher={Nature Publishing Group UK London},
  url={https://doi.org/10.1038/s41467-025-59716-z}
}

@article{Lowdin.1967.Rev.Mod.Phys.,
  title={Group algebra, convolution algebra, and applications to quantum mechanics},
  author={L{\"o}wdin, Per-Olov},
  journal={Rev. Mod. Phys.},
  volume={39},
  number={2},
  pages={259},
  year={1967},
  publisher={APS},
  url={https://doi.org/10.1103/RevModPhys.39.259}
}

@article{Hu.2015.J.Chem.Theory.Comput.,
  title={Excited-state geometry optimization with the density matrix renormalization group, as applied to polyenes},
  author={Hu, Weifeng and Chan, Garnet Kin-Lic},
  journal={J. Chem. Theory Comput.},
  volume={11},
  number={7},
  pages={3000--3009},
  year={2015},
  publisher={ACS Publications},
  url={https://doi.org/10.1021/acs.jctc.5b00174}
}

@article{Manna.2020.J.Chem.Phys.,
  title={Taming the excited states of butadiene, hexatriene, and octatetraene using state specific multireference perturbation theory with density functional theory orbitals},
  author={Manna, Shovan and Chaudhuri, Rajat K and Chattopadhyay, Sudip},
  journal={J. Chem. Phys.},
  volume={152},
  number={24},
  year={2020},
  publisher={AIP Publishing},
  url={https://doi.org/10.1063/5.0007198}
}

@article{Verma.2025.J.Chem.Theory.Comput.,
  title={Polynomial scaling localized active space unitary selective coupled cluster singles and doubles},
  author={Verma, Shreya and D’Cunha, Ruhee and Mitra, Abhishek and Hermes, Matthew and Gray, Stephen K and Otten, Matthew and Gagliardi, Laura},
  journal={J. Chem. Theory Comput.},
  volume={21},
  number={15},
  pages={7460--7470},
  year={2025},
  publisher={ACS Publications},
  url={https://doi.org/10.1021/acs.jctc.5c00745}
}

@book{Roberts.1965,
  title={Basic principles of organic chemistry},
  author={Roberts, John D and Caserio, Marjorie C},
  journal={W.A. Benjamin Inc.},
  year={1965},
  url={https://authors.library.caltech.edu/records/z1ms9-63w28}
}

@article{Hoffmann.1966.Tetrahedron,
  title={Extended h{\"u}ckel theory—v: Cumulenes, polyenes, polyacetylenes and cn},
  author={Hoffmann, Roald},
  journal={Tetrahedron},
  volume={22},
  number={2},
  pages={521--538},
  year={1966},
  publisher={Elsevier},
  url={https://doi.org/10.1016/0040-4020(66)80020-0}
}

@article{Diamond.2016.J.Mach.Learn.Res.,
  author       = {Steven Diamond and Stephen Boyd},
  title        = {{CVXPY}: A {P}ython-Embedded Modeling Language for Convex Optimization},
  journal      = {J. Mach. Learn. Res.},
  url          = {https://stanford.edu/~boyd/papers/pdf/cvxpy_paper.pdf},
  year         = {2016},
  url          = {https://dl.acm.org/doi/10.5555/2946645.3007036}
}

@article{Virtanen.2020.Nat.Methods,
  author  = {Virtanen, Pauli and Gommers, Ralf and Oliphant, Travis E. and
            Haberland, Matt and Reddy, Tyler and Cournapeau, David and
            Burovski, Evgeni and Peterson, Pearu and Weckesser, Warren and
            Bright, Jonathan and {van der Walt}, St{\'e}fan J. and
            Brett, Matthew and Wilson, Joshua and Millman, K. Jarrod and
            Mayorov, Nikolay and Nelson, Andrew R. J. and Jones, Eric and
            Kern, Robert and Larson, Eric and Carey, C J and
            Polat, {\.I}lhan and Feng, Yu and Moore, Eric W. and
            {VanderPlas}, Jake and Laxalde, Denis and Perktold, Josef and
            Cimrman, Robert and Henriksen, Ian and Quintero, E. A. and
            Harris, Charles R. and Archibald, Anne M. and
            Ribeiro, Ant{\^o}nio H. and Pedregosa, Fabian and
            {van Mulbregt}, Paul and {SciPy 1.0 Contributors}},
  title   = {{{SciPy} 1.0: Fundamental Algorithms for Scientific
            Computing in Python}},
  journal = {Nat. Methods},
  year    = {2020},
  volume  = {17},
  pages   = {261--272},
  adsurl  = {https://rdcu.be/b08Wh},
  url     = {https://doi.org/10.1038/s41592-019-0686-2},
}

@article{Sun.2020.J.Chem.Phys.,
  title={Recent developments in the PySCF program package},
  author={Sun, Qiming and Zhang, Xing and Banerjee, Samragni and Bao, Peng and Barbry, Marc and Blunt, Nick S and Bogdanov, Nikolay A and Booth, George H and Chen, Jia and Cui, Zhi-Hao and others},
  journal={J. Chem. Phys.},
  volume={153},
  number={2},
  year={2020},
  publisher={AIP Publishing},
  url={https://doi.org/10.1063/5.0006074}
}

@article{McClean.2020.Quantum.Sci.Technol.,
  title={OpenFermion: the electronic structure package for quantum computers},
  author={McClean, Jarrod R and Rubin, Nicholas C and Sung, Kevin J and Kivlichan, Ian D and Bonet-Monroig, Xavier and Cao, Yudong and Dai, Chengyu and Fried, E Schuyler and Gidney, Craig and Gimby, Brendan and others},
  journal={Quantum Sci. Technol.},
  volume={5},
  number={3},
  pages={034014},
  year={2020},
  publisher={IOP Publishing},
  url={https://doi.org/10.1088/2058-9565/ab8ebc}
}

@article{Goings.2022.Proc.Natl.Acad.Sci.U.S.A.,
  title={Reliably assessing the electronic structure of cytochrome P450 on today’s classical computers and tomorrow’s quantum computers},
  author={Goings, Joshua J and White, Alec and Lee, Joonho and Tautermann, Christofer S and Degroote, Matthias and Gidney, Craig and Shiozaki, Toru and Babbush, Ryan and Rubin, Nicholas C},
  journal={Proc. Natl. Acad. Sci. U. S. A.},
  volume={119},
  number={38},
  pages={e2203533119},
  year={2022},
  publisher={National Academy of Sciences},
  url={https://doi.org/10.1073/pnas.2203533119}
}

@article{Low.2025.Phys.Rev.X,
  title={Fast quantum simulation of electronic structure by spectral amplification},
  author={Low, Guang Hao and King, Robbie and Berry, Dominic W and Han, Qiushi and DePrince III, A Eugene and White, Alec F and Babbush, Ryan and Somma, Rolando D and Rubin, Nicholas C},
  journal={Phys. Rev. X},
  volume={15},
  number={4},
  pages={041016},
  year={2025},
  publisher={APS},
  url={https://doi.org/10.1103/pb2g-j9cw}
}

@article{Reiher.2017.Proc.Natl.Acad.Sci.U.S.A.,
  title={Elucidating reaction mechanisms on quantum computers},
  author={Reiher, Markus and Wiebe, Nathan and Svore, Krysta M and Wecker, Dave and Troyer, Matthias},
  journal={Proc. Natl. Acad. Sci. U. S. A.s},
  volume={114},
  number={29},
  pages={7555--7560},
  year={2017},
  publisher={National Academy of Sciences},
  url={https://doi.org/10.1073/pnas.1619152114}
}

\onecolumn\newpage
\appendix
\section{Figures} \label{s:appendix-figures}

\begin{figure}[ht]
    \centering
    \includegraphics[width=0.8\linewidth, trim={0 2.25cm 0 2cm}]{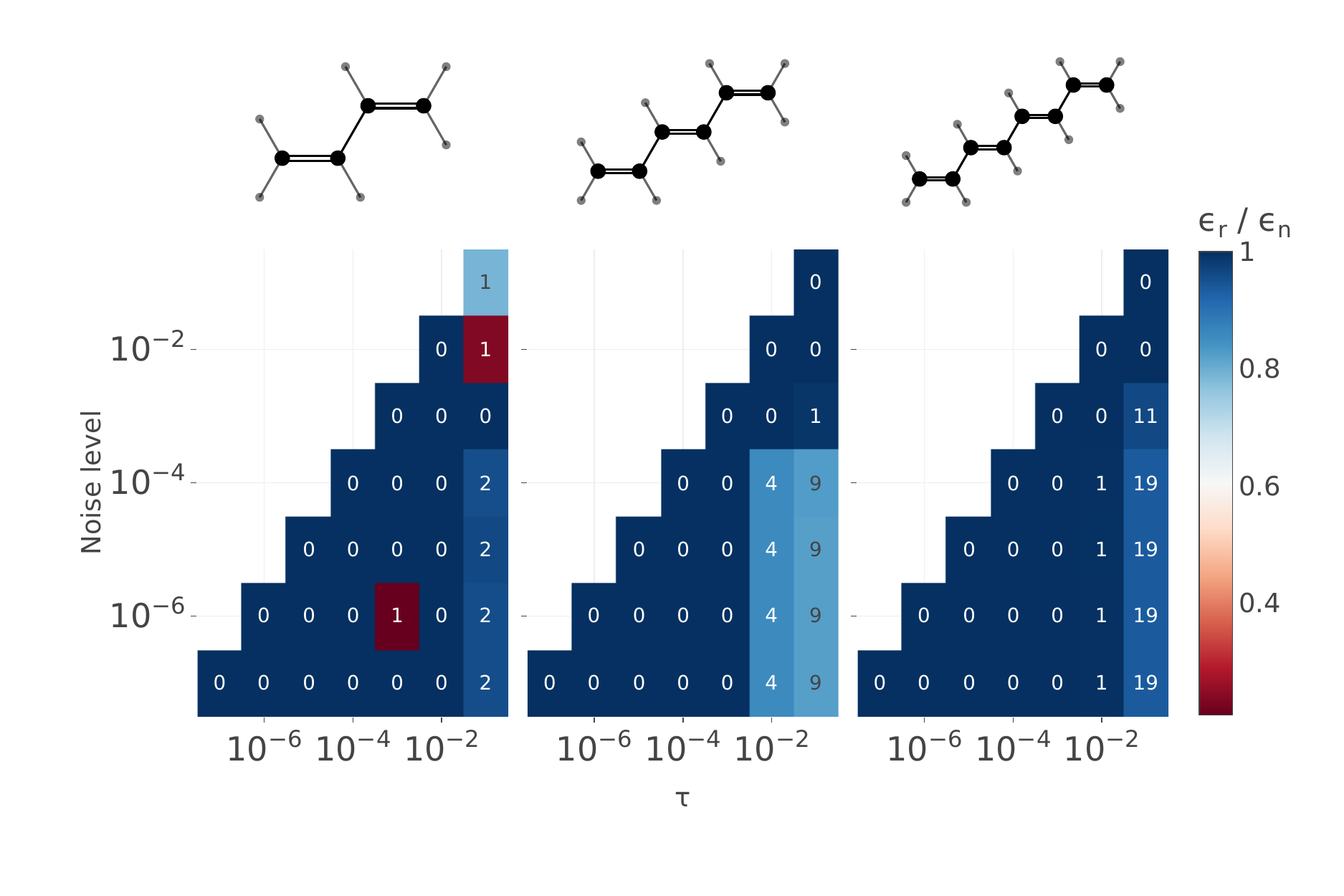}
    \caption{Ratio of the absolute error with respect to the CASSCF solution made by the rotation thresholding approach with rotation angle $\theta_{\text{heuristic}}$ ($\epsilon_r$) and the same error made by naively thresholding ($\epsilon_n$), calculated for in length increasing all-trans polyene chains. Noise level indicates the variance of the artificially applied Gaussian noise, and $\tau$ is the thresholding parameter. Numbers in the heatmap represent the amount of additional states kept by the rotation approach, that were discarded when applying naive thresholding. Convergence scaling factor $\gamma=1.0$.}
    \label{fig:Polyenes-heuristic-grid}
\end{figure}

\begin{figure}[ht]
    \centering
    \includegraphics[width=0.8\linewidth, trim={0 2cm 0 2cm}]{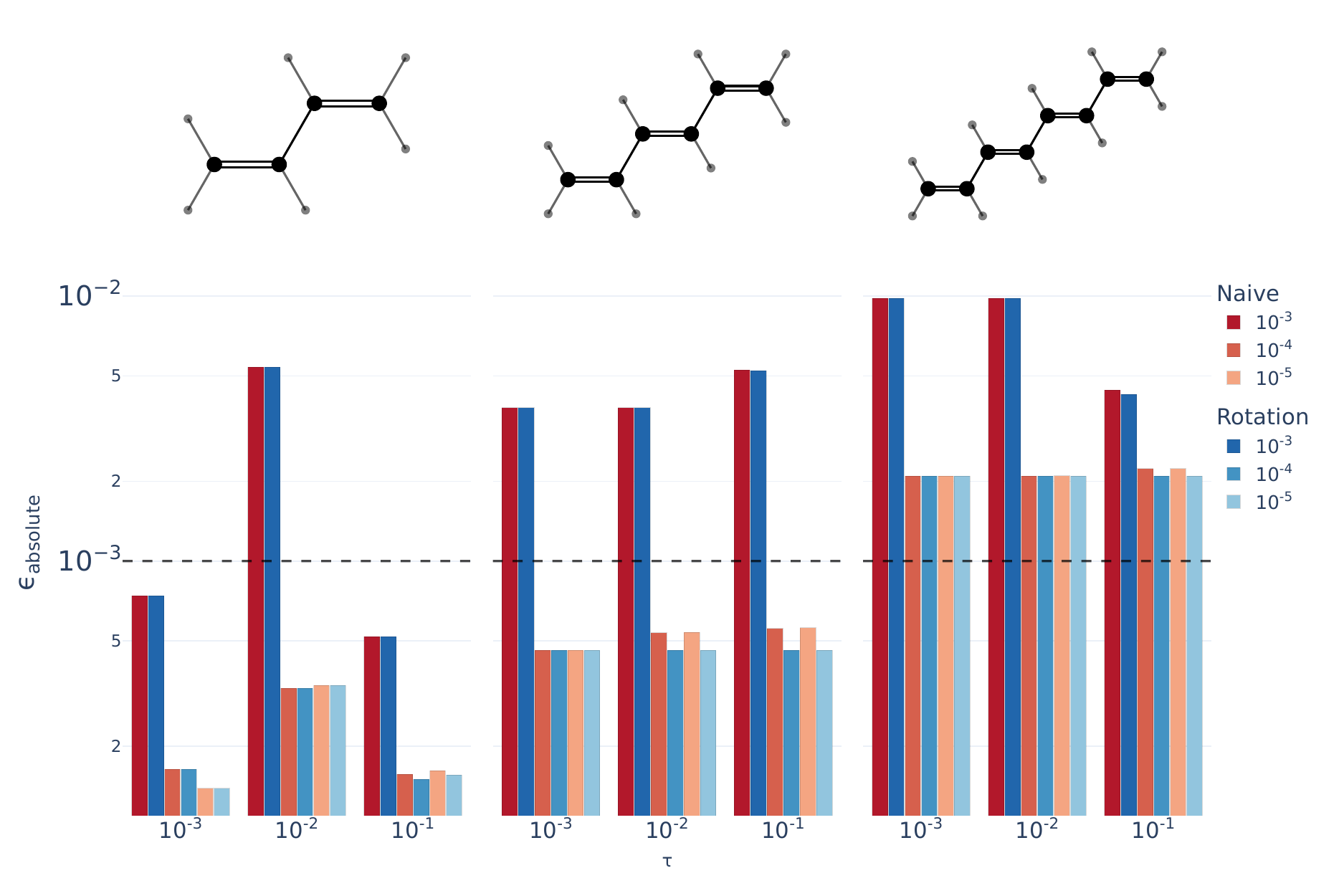}
    \caption{Absolute error ($\epsilon_\text{absolute}$) in log-scale with respect to the CASSCF solution made by the rotation thresholding approach with rotation angle $\theta_{\text{heuristic}}$ (blue bars) and the same error made by naively thresholding (red bars) for in length increasing all-trans polyene chains and three different noise levels, at all thresholding parameters $\tau \geq$ noise level. Dashed line represents chemical accuracy. Convergence scaling factor $\gamma=1.0$.}
    \label{fig:Polyenes-heuristic-bars}
\end{figure}

\begin{figure}[ht]
    \centering
    \includegraphics[width=0.8\linewidth, trim={0 2cm 0 1cm}]{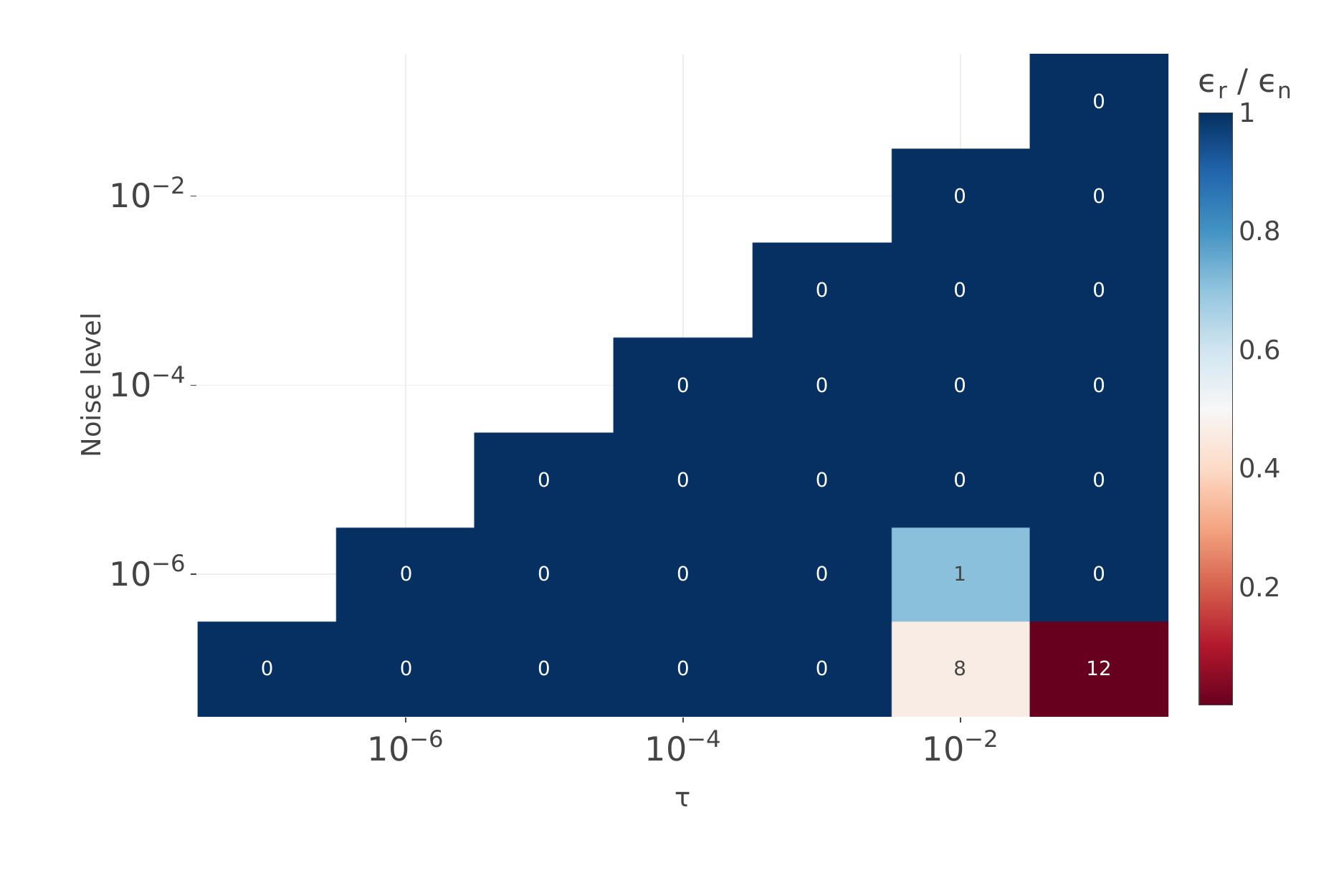}
    \caption{Ratio of the absolute error with respect to the CASSCF solution made by the rotation thresholding approach with rotation angle $\theta_{\text{heuristic}}$ ($\epsilon_r$) and the same error made by naively thresholding ($\epsilon_n$), calculated for the Fe(III)-NTA chelate complex. Noise level indicates the variance of the artificially applied Gaussian noise, and $\tau$ is the thresholding parameter. Numbers in the heatmap represent the amount of additional states kept by the rotation approach, that were discarded when applying naive thresholding. Convergence scaling factor $\gamma=0.75$.}
    \label{fig:FeNTA-heuristic-grid}
\end{figure}

\begin{figure}[ht]
    \centering
    \includegraphics[width=0.8\linewidth, trim={0 2cm 0 2cm}]{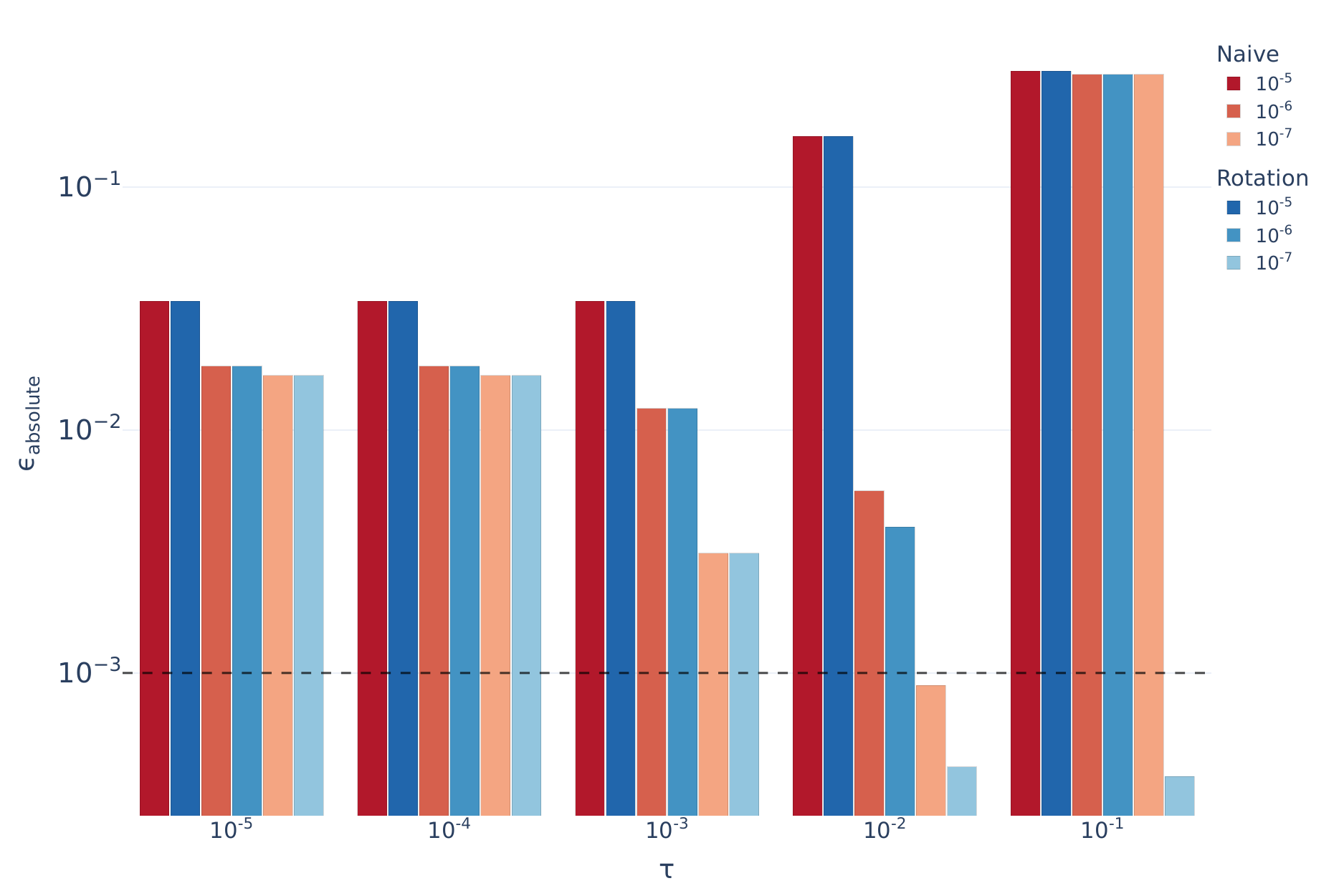}
    \caption{Absolute error ($\epsilon_\text{absolute}$) in log-scale with respect to the CASSCF solution made by the rotation thresholding approach with rotation angle $\theta_{\text{heuristic}}$ (blue bars) and the same error made by naively thresholding (red bars) for three different noise levels, at all thresholding parameters $\tau \geq$ noise level, calculated for the Fe(III)-NTA chelate complex. Dashed line represents chemical accuracy. Convergence scaling factor $\gamma=0.75$.}
    \label{fig:FeNTA-heuristic-bars}
\end{figure}

\end{document}